\documentclass[11pt]{article}%{amsart}%{article}
\usepackage{amssymb,amsfonts,amsmath,mathrsfs,nccmath,geometry,authblk}
\usepackage{graphicx,xcolor,pstricks}
\usepackage[T1]{fontenc}
\usepackage{parskip}
\numberwithin{equation}{section}
\makeatletter
\let\@fnsymbol\@zero
\newcommand{\sech} { \mathop{\rm sech}}
\newcommand{\hpsi} {\widehat{\psi}}
\newcommand{\nvarphi} {\varphi^{\rm st}}
\newcommand{\nphi} {\phi^{\rm st}}
\newcommand{\npsi} {\psi^{\rm st}}
\begin{document}
 
\title{Stability of solitary waves in nonlinear Klein-Gordon  equations}

\author[1]{Pablo Rabán\thanks{emails: \texttt{pabloraban@outlook.com, ran@us.es, niurka@us.es}}}
\affil[1]{Departamento de Física Teórica, Facultad de Ciencias Físicas, Universidad Complutense de Madrid, Plaza de Ciencias 1, Ciudad Universitaria,	28040, Madrid, Spain}
\author[2]{Renato Alvarez-Nodarse}
\affil[2]{IMUS, Universidad de Sevilla, Departamento de An\'alisis Matem\'atico, Universidad de
	Sevilla, c/Tarfia s/n, 41012, Sevilla, Spain}
\author[3]{Niurka R.\ Quintero}
\affil[3]{F\'\i sica Aplicada I, Escuela Polit\'ecnica Superior, Universidad de
	Sevilla, Virgen de \'Africa 7, 41011, Sevilla, Spain}

 \date{ \today}
 
 \maketitle

%%%%%%%%%%%%%%%%%%%%%%%%%%%%%%%%%%%%%%%%%%%%%%%%%%%%%%%%%%%%%%%%%%%%%%%%%%%%%%%%%%
\begin{abstract}
The stability of topological solitary waves and pulses in  one-dimensional nonlinear Klein-Gordon systems is revisited.
The linearized equation describing small deviations around the static solution  
leads to a Sturm-Liouville problem, which is solved in a systematic way for the 
$-l\,(l+1)\,\sech^2(x)$-potential, showing the orthogonality and completeness relations 
fulfilled by the set of its solutions for all values $l\in\mathbb{N}$. 
This approach enables the linear stability of kinks and pulses of certain nonlinear Klein-Gordon equations to be determined. The inverse problem, which starts from Sturm-Liouville problem and obtains nonlinear Klein-Gordon potentials,  is also revisited and solved in a direct way. The exact solutions (kinks and pulses) 
for these potentials are calculated, even when the nonlinear potential is not explicitly known. 
The kinks are found to be stable, whereas the pulses are unstable. The stability of 
the pulses is achieved by introducing certain spatial inhomogeneities.
\end{abstract}
%%%%%%%%%%%%%%%%%%%%%%%%%%%%%%%%%%%%%%%%%%%%%%%%%%%%%%%%%%%%%%%%%%%%%%%%%%%%%%%%%%
 
\section{Introduction\label{intro}}

Nonlinear Klein-Gordon equations model a plethora of phenomena such as the
existence of bound oscillatory states and resonance windows in the kink-antikink interaction \cite{getmanov:1976,makhankov:1978,sugiyama:1979,aubry:1976,campbell:1983} in the presence and in the absence of internal modes \cite{campbell:1986,dorey:2011,demirkaya:2017,campbell:2019}, the fading of the kink’s wobbling due to the second-harmonic radiation \cite{barashenkov:2009,barashenkov:2019}, the phase transitions in the Ginzburg-Landau theory \cite{landau:1937,landau:1950,khare:2014}, the motion of domain walls \cite{buijnsters:2014}, and the existence of kinks with power-law tail asymptotics that
give rise to long-range interactions in the even-higher-order field theories \cite{christov:2019,saxena:2019}.

In a one-dimensional system, the Hamiltonian corresponding to the nonlinear Klein-Gordon equation is a functional of the field $\varphi(x,t)$, defined 
in the following way: 
\begin{equation}\label{H-E}
H[\varphi(x,t)]=\int_{\mathbb{R}}\,dx\, 
\left\{
\frac{\varphi_{t}^2(x,t)}2+\frac{\varphi_{x}^2(x,t)}2+U[\varphi(x,t)]
\right\},
\end{equation}
where $U[\varphi(x,t)]$ is the nonlinear Klein-Gordon potential, the integral is performed across the whole space 
$x \in \mathbb{R}$, and $t$ and $x$ subscripts henceforth denote the partial derivative with respect 
to time and position, respectively. Since the energy of the system must be finite, the spatial derivative of the 
field, $\varphi_{x}(x,t)$, should be a bounded function for all $x  \in \mathbb{R}$ and $t\geq0$. 

From the Hamiltonian field equations \cite{goldstein:1980}, the one-dimensional nonlinear Klein-Gordon system is given by   
\begin{equation}
	\label{sg}
\varphi_{tt}(x,t)-\varphi_{xx}(x,t)=- \frac{d U}{d \varphi}[\varphi(x,t)], 
\end{equation}
where the function $\varphi(x,t) \in C^2_{\infty}(\mathbb{R}\times[0,+\infty))$. Here,     
$C^2_{\infty}(\Omega)$ denotes the space of functions $f:\Omega\mapsto\mathbb{C}$ 
that are two times differentiable with continuous second order
partial derivatives in $\Omega$, and such that $f$ and their first partial derivatives are bounded in $\Omega$. 
Throughout this study it is assumed that the nonlinear Klein-Gordon potential $U[\varphi(x,t)]$  
has at least two extrema to guarantee  the existence of kink or pulse-like solutions. 

In particular, a pulse may emerge when  the nonlinear potential $U[\varphi(x)]$ has a minimum and a 
maximum as two consecutive extrema, while the appearance of a topological wave called kink requires 
that $U[\varphi(x)]$ has two  consecutive minima sharing  the same value. Hereafter the function $\nphi_p(x)$ 
represents the static pulses  and $\nvarphi(x)$, the kinks at rest.

The dynamics and stability of the \textit{sech}-shape solution of certain cubic potential  were 
investigated in Ref. \cite{fullin:1978}, where the pulse was found to be unstable. 
Moreover,  kink solutions $\nvarphi(x)$ have been derived  
for the sine-Gordon potential \cite{scott:1969a}, the double sine-Gordon potential \cite{condat:1983},
and the $\varphi^n$ ($n= 4, 6, 8$) potentials \cite{dashen:1974,lohe:1979} (see also the recent reviews 
\cite{dauxois:2006,malomed:2014,campbell:2019,saxena:2019} wherein developments of higher-order
field theories are discussed). Since Eq.\ \eqref{sg} is  Lorentz invariant,  
traveling solutions can be obtained from the static solutions by a boost transformation.

At the boundaries, the topological waves verify   
\begin{eqnarray}
	\label{sgbc2}
	\lim_{x \to + \infty}\varphi(x,t)&=&\lim_{x \to - \infty}\varphi(x,t) \pm Q, 
\end{eqnarray}
where the signs $\pm$ refer to the kink and antikink solution, respectively, and the constant $Q > 0$ is the so-called topological charge. The pulses also satisfy Eq.\ \eqref{sgbc2}  setting $Q=0$.

As a matter of fact, the equations that govern physical systems, such as the propagation of magnetic flux along the Josephson junctions \cite{barone:1971}, the dynamics of the azimuthal angle of the unit vector of
	magnetization of ferromagnetic materials \cite{barashenkov:1991}, the resonant soliton-impurity interactions 
	\cite{kivshar:1991}, the unidirectional motion of kinks due to zero-average forces \cite{salerno:2002,morales-molina:2003,ustinov:2004,quintero:2014}, 
	the stabilization of wobbling kinks \cite{barashenkov:2019}, and the traveling of dislocations  along the colloids \cite{casic:2013}, are modeled by nonlinear Klein-Gordon equations with external and parametric forces and damping. Therefore, the observation of these waves in nature 
	and in experiments depends on their stability \cite{scott:1999}.    
From the study of  stability it can be determined whether the perturbed solution of  Eq.\ (\ref{sg}) does not deviate 
too far from the exact solution when the perturbations are small enough, that is, whether the exact solution would be detected in a real system.

The stability of the sine-Gordon waves was studied by 
Scott more than 50  years ago, first by means of an average Lagrange method \cite{scott:1969a}, and second by 
employing a more accurate technique, namely the eigenfunction expansion \cite{scott:1969}, introduced by Parmentier 
to study the stability of a nonlinear transmission line \cite{parmentier:1967}. As a result of the second methodology, a  Sturm-Liouville problem  
was obtained and discussed, although not completely solved. Its spectrum has both a discrete and a continuous part \cite{rubinstein:1970}. 
Indeed, since Eq.\ (\ref{sg}) is translationally invariant, there is always a zero mode associated  
to the zero eigenvalue (zero frequency) and to an eigenfunction proportional to the spatial derivative of the 
field  \cite{buratti:1968,scott:1969}. The continuous spectrum exists due to the infinite domain \cite{morse2:1953}.

The exact analytical solution of the Sturm-Liouville problem was found 
for the sine-Gordon kink, the $\varphi^4$ kink, the $\varphi^6$ kink, and also for the  pulse 
	of certain cubic potential     
\cite{rubinstein:1970,dashen:1974,fullin:1978,lohe:1979}, among others. 
However, in some cases, only numerical 
solutions of this problem have been found either because the solution of Eq. (\ref{sg}) is implicit, as in 
the $\varphi^8$  equation  \cite{lohe:1979}, or because it has been impossible to analytically solve the 
corresponding Sturm-Liouville problem, as in the double sine-Gordon system \cite{campbell:1986}.

The main goal of the current investigation is to revise the Sturm-Liouville problem 
equivalent to the one-dimensional Schr\"odinger  equation  with the 
 P\"oschl-Teller potential \cite{poschl:1933}, 
$-l\,(l+1)\,\sech^{2}(x)$, $l\in\mathbb{N}$, which is associated to the stability of nonlinear waves of 
several aforementioned nonlinear Klein-Gordon potentials. This potential is one of the most 
useful potentials in mathematical physics. It appears in Optics \cite{epstein:1930}, in Quantum Mechanics \cite{eck:1930} 
(see also page 768 in Ref. \cite{morse:1953}, page 94 in Ref. \cite{flugge:1994}, and  
page 73 in Ref.  \cite{landau:1977}), in the $N$-soliton
solution of the Korteweg and de-Vries equation \cite{drazin:1989}, and in the stability study of certain static solutions \cite{yang:2000}. 
The above potential, belongs to the special class of potentials for which the one-dimensional Schr\"odinger  equation 
can be exactly solved in terms of special functions (see page 768 in Ref.\ \cite{morse:1953}).

The inverse problem, proposed by Christ and Lee in Ref.  \cite{christ:1975} for the specific case of the kink solutions, investigates the existence of  other nonlinear Klein-Gordon kinks or pulses whose stability is associated with the  
P\"oschl-Teller potential?   
 They started from the translational mode corresponding to the P\"oschl-Teller potential,  and  
	partially, although not explicitly,  constructed the sine-Gordon ($l=1$) and $\varphi^4$ ($l=2$) potentials. The explicit construction of the potential $U(\varphi)$ by solving the resulting differential equation for $U(\varphi)$ is due to 
	Trullinger and Flesch \cite{trullinger:1987}. They obtained two solutions for $U(\varphi)$, one for the odd values of $l$ and the other for the even values of $l$, and they found that these potentials can be expressed in terms of the Student's $t$-distribution of probability theory.

Despite all these studies, to the best of our knowledge, 
there is no rigorous analysis of the corresponding Sturm-Liouville problem, 
nor a detailed proof of the orthogonality and completeness of its  eigenfunctions for all values of $l\in\mathbb{N}$, nor a solution of the inverse problem in a direct way. It is the aim of the current study  to complete the aforementioned studies.

Section \ref{sec2} provides the outline of the linear stability analysis of the static solution, 
either $\nvarphi(x)$ or $\nphi_p(x)$, of Eq.\ (\ref{sg}), and derives the associated Sturm-Liouville problem. 
This Section ends with a precise definition of the stability, which requires the positiveness 
of all the eigenvalues (squared eigenfrequencies).  
This definition is a consequence of the ansatz employed, in order to solve the Sturm-Liouville problem. The subsequent Section \ref{sec3} solves the Sturm-Liouville problem with the  potential, $-l\,(l+1)\,\sech^{2}(x)$, $l\in\mathbb{N}$, in a systematic way, including a detailed proof of the orthogonality and completeness relations. This is a very crucial result, since 
in practical applications the spatial component of the solution of certain perturbed nonlinear 
Klein-Gordon equations is written as an expansion in the set  of eigenfunctions (see e.g. 
\S5.2 on page 144   %\cite[\S5.2 page 144]{dauxois:2006}. 
in Ref. \cite{dauxois:2006}). 
Specifying the values of $l=1$ and $l=2$, it is shown that the  sine-Gordon and $\varphi^4$ kinks, respectively, are stable. The values of $l=3$ and $l=2$ are related with the unstable pulses of the 
cubic and quartic potentials, respectively.

Section \ref{sec4} addresses this issue and reconstructs the theory in a similar way to that in  Ref. 
\cite{trullinger:1987}; however,  the problem is solved in a more direct way, without using the Student's $t$-distribution. Our procedure has two advantages with respect to the previous analyses of Refs. \cite{christ:1975,trullinger:1987}. First, our analysis is valid for all values of $l$ and the solution of the second-order differential equation for $U(\varphi)$ is represented in a closed form in terms of  the  hypergeometric function, where $l$ is a parameter. Second, all kink solutions can be obtained by a recurrence relation, where the sequence of kinks depends on the value of $l$. 
Section \ref{sec4} obtains two  families of nonlinear Klein-Gordon potentials such that the P\"oschl-Teller potential appears 
in their corresponding Sturm-Liouville problems. For the first family, the exact analytical kinks  are obtained. It is demonstrated that all kinks are stable. 
For the second family, the pulses are derived. Although all the pulses found are unstable, Section \ref{sec5} provides guidelines for their stabilization 
through inhomogeneous forces. Finally, Section \ref{sec6} discusses our main results and draws general conclusions.

\section{The nonlinear Klein-Gordon equation and its corresponding Sturm-Liouville problem} \label{sec2}

Due to the Lorentz invariance of Eq.\ (\ref{sg}), it is sufficient to investigate the stability of 
the static kink \cite{dauxois:2006},  $\varphi(x,t)=\nvarphi(x)$, which satisfies the following  equation  
\begin{equation}
\label{sg1}
\nvarphi_{xx}(x)= \frac{d U}{d \varphi}[\nvarphi(x)],
\end{equation} 
where the nonlinear Klein-Gordon potential $U$ has at least two local minima, 
which are reached by $U[\nvarphi(x)]$ as $x\to\pm\infty$. 
Notice that, Eq.\ \eqref{sg1} resembles the second Newton law for a particle in a potential $-U[\nvarphi(x)]$ \cite{christ:1975}. 
This is the reason why $-U$ is called a pseudo-potential \cite{hansen:1979}. 
Within this framework, the variables $x$ and $\nvarphi(x)$ play the role of time and position, respectively.

\begin{figure}[h]
	\begin{tabular}{cc}
		\includegraphics[width=0.5\linewidth]{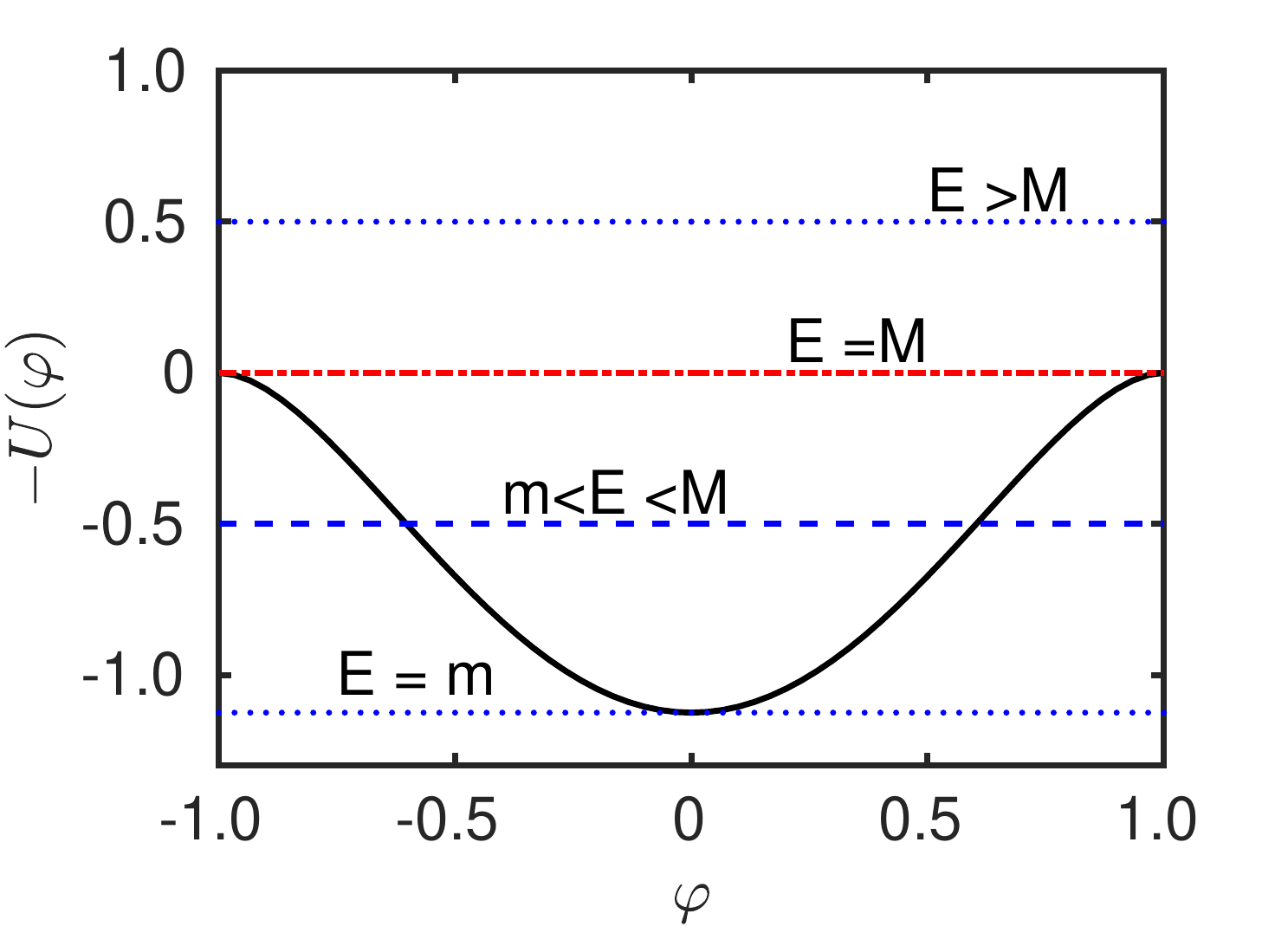} &
		\includegraphics[width=0.5\linewidth]{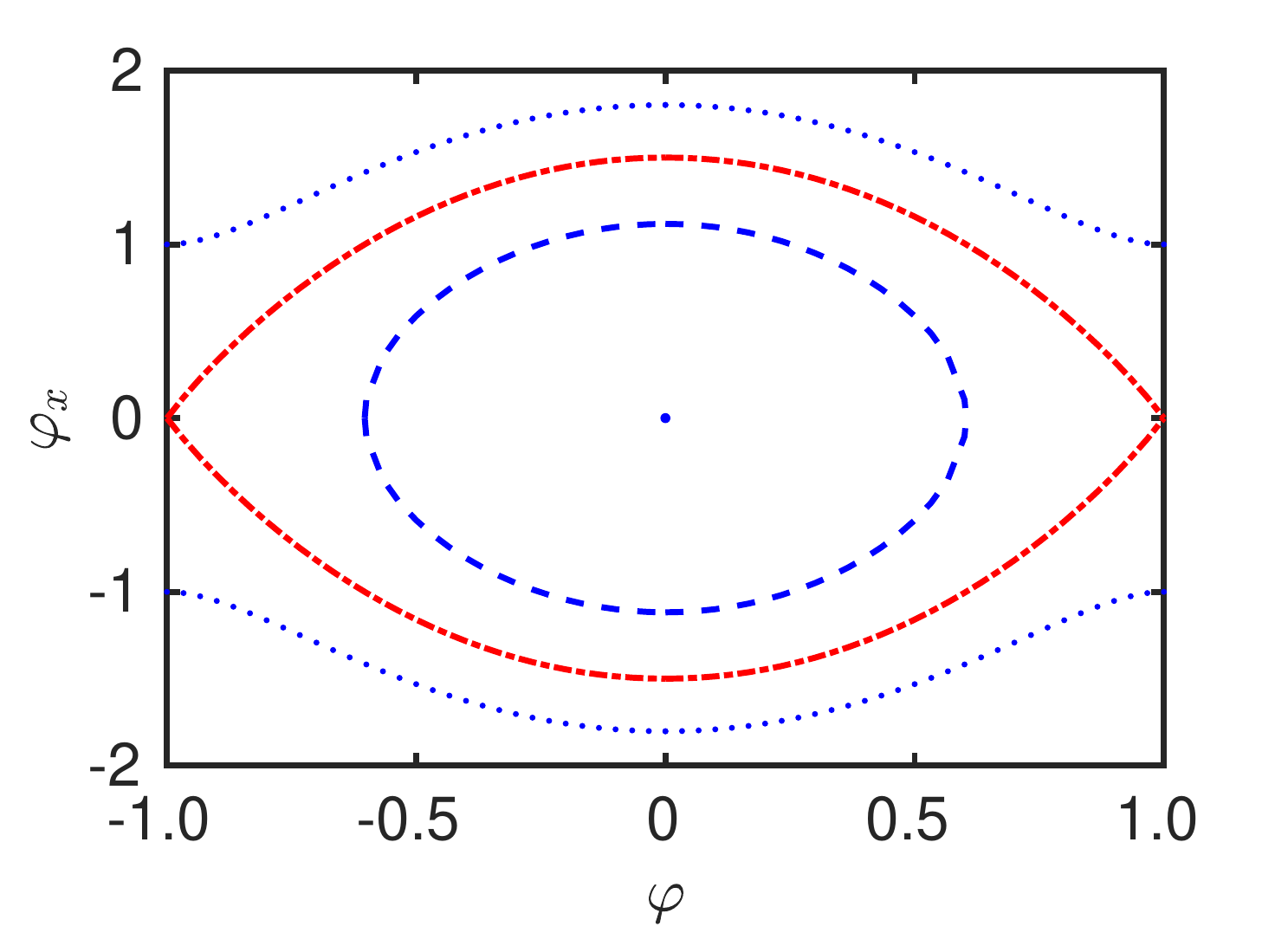}
	\end{tabular}
	\caption{Left-hand panel: the pseudo-potential $-U(\varphi)$ is shown (solid black line), and four  values of energy related with different types of motions (see the text): $E=m$  and $E>M$ 
		(dotted blue lines), $m<E<M$ (dashed blue line), and $E=M$ (dot-dashed red line).
		Right-hand panel: the separatrix, represented with a dot-dashed red line ($E=M$), it separates the region where the motion is unbounded (dotted blue line, $E>M$) from the region where the oscillatory motion takes place (dashed blue line, $m<E<M$). The particle is at rest at the origin.}	
	\label{fig1}
\end{figure}

By integrating Eq.\ (\ref{sg1}), the first integral of motion reads 
\begin{equation}
	\label{sg2}
	E=\frac{(\nvarphi_{x})^2}{2}-U[\nvarphi(x)],
\end{equation}    
where $E$ denotes the total energy of the Newtonian particle, and $(\nvarphi_{x})^2/2$ its  kinetic energy. 
Three different cases are distinguished according to the value of $E$. 
When the energy $E>M$, where $M$ is the maximum of the pseudo-potential, the particle is always moving, similar to the rotatory motion of the simple 
	pendulum, see Fig.\ \ref{fig1}. If $m\le E<M$, where  $m$ is the minimum of the pseudo-potential, the particle oscillates except for $E=m$ when the particle remains at rest. The separatrix at $E=M$ separates oscillatory and rotatory motions, see the phase portrait in Fig. \ref{fig1}.  
The kink (antikink), $\nvarphi(x)$, is represented precisely by the separatrix of the dynamical system (\ref{sg1}), which connects two maxima of the pseudo-potential $-U[\nvarphi(x)]$, that is, two minima of the potential  $U[\nvarphi(x)]$, when $x \to \pm \infty$. As a consequence,
\begin{eqnarray}
	\label{sg2lim2}
	&&\lim_{x \to \pm \infty} \frac{dU}{d\varphi}[\nvarphi(x)]=0, \\
	\label{sg2lim3}
	&&\lim_{x \to \pm \infty} \frac{d^2U}{d\varphi^2}[\nvarphi(x)] \ge 0.
\end{eqnarray}
Without any loss of generality, it is assumed that the minimum of the potential is reached at zero, that is, 
\begin{eqnarray}
	\label{sg2lim}
	&&\lim_{x \to \pm \infty} U[\nvarphi(x)]=0.
\end{eqnarray}
This implies that $U[\nvarphi(x)] \ge 0$ between the two minima. Furthermore, it sets $E=0$ in  (\ref{sg2}). 
Notice that these conditions on the potential and its derivatives are also satisfied by a static pulse, solution of 
Eq.\ \eqref{sg1}. Indeed, a pulse lies on the separatrix that begins and ends at the same equilibrium point, which is a minimum of the potential. 

From Eq.\ (\ref{sg2}), it follows that the static kink, or static pulse, can be calculated by performing the following integral 
\begin{equation}
	\label{solu1}
	\int \frac{d\varphi}{\sqrt{2\,U(\varphi)}}= x+C,
\end{equation} 
where the constant $C$ is set to zero  
due to the translational invariance. By considering the sine-Gordon potential 
\begin{equation}
	\label{sgp}
	U(\varphi)=1-\cos(\varphi) 
\end{equation}
in Eq. \eqref{solu1}, and by integration, the static kink has the form 
\begin{equation}
	\label{sgk}
	\nvarphi(x)=4\,\arctan[\exp(x)].
\end{equation} 
By applying a similar procedure with the  $\varphi^4$ potential
\begin{equation}
	\label{phi4p}
	U(\varphi)=\frac{(1-\varphi^2)^2}{2},
\end{equation}
we obtain the static kink
\begin{equation}
	\label{phi4k}
	\nvarphi(x)=\tanh(x).
\end{equation}  
By integrating Eq.\ \eqref{solu1} with the cubic and quartic potentials  shown in the first column of Table \ref{tabla2}, 
the static pulses are calculated (second column of Table \ref{tabla2}). 
The second-order differential equation \eqref{sg1} with the cubic and quartic potentials also appears 
	when we find soliton solutions in the KdV equation \cite{korteweg:1985,drazin:1989} and in the nonlinear Schr\"odinger (NLS) equation \cite{chiao:1964,scott:1999,sulem:1999}, respectively. The KdV soliton is non-topological and has the same shape as the pulse of the cubic potential, whereas the envelope part of the NLS soliton and the pulse of the quartic potential have the same shape.

\begin{table}[ht!]
	\begin{center}
		\begin{small}
			{\renewcommand{\arraystretch}{1.5}%		
				\begin{tabular}{|c|c|c|c|c|}
					\hline		
					$U(\phi)$                  &   $\nphi_p(x)$    & $\npsi(x)$         &    $V(x)$  & $\omega_{ph}$\\
					\hline	\hline
					$2\,\phi^2\,(1-\phi)$             &   $\frac{1}{ \,\cosh^2(x)}$   & $\frac{-2\,\tanh(x)}{\cosh^2(x)}$ & $-\frac{12}{\cosh^2(x)}$  & $2$ \\
					\hline
					$\frac{\phi^2}{2}\,(1-\phi^2)$  & $\frac{1}{ \,\cosh(x)}$    &  $\frac{-\tanh(x)}{ \,\cosh(x)}$  &   $-\frac{6}{\cosh^2(x)}$  & $1$ \\ 
					\hline
			\end{tabular}}
		\end{small}%
	\caption{For the   cubic and quartic potentials, $U(\phi)$, 
		 the unstable solitary waves (pulses) $\nphi_p(x)$, 
	the eigenfunction $\npsi(x)=d\nphi_{p}/dx$ associated to the zero frequency, 
	the potential $V(x)$ \eqref{utildep}, and the lowest frequency of the continuous spectrum, $\omega_{ph}$  are shown.}
		\label{tabla2}
	\end{center}
\end{table}

In order to discuss the stability of the static kink, $\nvarphi(x)$, Eq.\ (\ref{sg}) is linearized around $\nvarphi(x)$, 
that is, the function \cite{parmentier:1967,scott:1969} 
\begin{equation}
	\label{sg5}
	\varphi(x,t)= \nvarphi(x)+\Psi(x,t), 
\end{equation}
is introduced in Eq. (\ref{sg}). This implies that the function  $\Psi(x,t)$ satisfies the following 
linear wave equation with a source term
\begin{equation}
\label{sg6}
 \Psi_{tt}(x,t)-\Psi_{xx}(x,t)=-U''[\nvarphi(x)] \Psi(x,t), 
\end{equation}
where the prime denotes the derivative of $U[\varphi(x)]$ with respect to $\varphi(x)$. 
It is important to bear in mind that in the above relation (\ref{sg5}), the 
second term should be small in comparison with $\nvarphi(x)$.  
This can be achieved 
if the $L^{\infty}$-norm of $\Psi(x,t)$ is finite for all $t \ge t_0$, where $t_0$ is 
the initial time, that is, $\sup_{x\in\mathbb{R},t\geq t_0}|\Psi(x,t)|< +\infty$,
and sufficiently small in comparison with $\sup_{x\in\mathbb{R}}|\nvarphi(x)|$. 
We recall that, since the energy of the system must be finite, $\varphi_x(x,t)$ should be a bounded function in 
$x \in \mathbb{R}$, then so should $\Psi_x(x,t)$.
 
The solution of Eq.\ (\ref{sg6}) is represented by the following ansatz  \cite{parmentier:1967,scott:1969,saxena:2019}
\begin{equation} \label{ansatz}
	\Psi(x,t)= (c_1 e^{i\omega t}+c_2 e^{-i\omega t})\,\psi(x),
\end{equation}
where $\psi(x)$ is a complex function  and the complex constants $c_1$ and $c_2$ are chosen such that $\Psi(x,t)\in\mathbb{R}$. 
By inserting Eq.\ (\ref{ansatz}) into Eq.\ (\ref{sg6}),  it is straightforwardly deduced that 
$\psi(x)$ verifies the following  Sturm-Liouville problem 
\begin{equation}
	\label{sg8}
	 \psi_{xx}(x)+\left(\omega^2-U''[\nvarphi(x)]\right)\psi(x)=0, 
\end{equation}
where $\omega^2$ (squared eigenfrequencies) are the eigenvalues  and it is required that $\psi(x)$ as well as its first derivative $\psi_{x}(x)$ are bounded 
and	continuous functions on $\mathbb{R}$ (recall that the energy of the system must be finite). 
Notice that Eq.\ (\ref{sg8}) can be written as $L \psi=\omega^2 \psi$, where the operator $L=-d^2/dx^2+U''$ is self-adjoint, therefore all their eigenvalues $\omega^2$ are real \cite{takh:2008}.
     
The Sturm-Liouville problem  (\ref{sg8}) is solved when the set of infinite (for infinite
domain) eigenfunctions $\{\psi(x)\}$ with their corresponding real eigenvalues 
 $\{\lambda=\omega^2\}$  
are found. The spectrum contains a set of discrete eigenvalues and the
so-called continuous spectrum. 
Useful features that satisfy the real eigenvalues of Eq.\ (\ref{sg8})  include: 
(i) the $N+1$ discrete eigenvalues form a continuously increasing sequence of real numbers bounded from below 
$\lambda_0<\lambda_1<\lambda_2<\cdots, \lambda_{N}$,  such that $\lambda_{i+1}>\lambda_{i}$, $i=0, 1, \cdots, N-1$;   
(ii) If $\psi_{i+1}(x)$ and $\psi_{i}(x)$ are the  eigenfunctions
associated to the discrete eigenvalues  $\lambda_{i+1}$ and $\lambda_{i}$, respectively, then $\psi_{i+1}(x)$
has one  more zero than does $\psi_{i}(x)$. As a consequence, the   eigenfunction $\psi_0(x)$ corresponding 
to $\lambda_0$ has the least possible number of zeros \cite{morse:1953};   
(iii) The proof of statement (ii), given in Ref. \cite{morse:1953}, 
can be generalized to the continuous spectrum and it can be shown that, 
in general, given two eigenvalues $\lambda_a<\lambda_b$, if $\psi_{a}(x)$ 
and $\psi_{b}(x)$ are their corresponding eigenfunctions, then $\psi_a(x)$ has 
no more zeros than does $\psi_{b}(x)$. In fact, this is also true for any two 
eigenvalues,  independently of whether
they belong to the continuous or discrete spectra.
Therefore, if an eigenfunction has no zeros, its corresponding eigenvalue is the 
lowest.

Another useful property of the  Sturm-Liouville problem  (\ref{sg8}) is related with
the zero mode, that is, the eigenfunction associated to $\omega=0$.  
Since the function $\nvarphi(x)$ is the solution of Eq.\ (\ref{sg1}), 
its derivative satisfies  $(\nvarphi_{x})_{xx}-U''[\nvarphi(x)] \nvarphi_{x}=0$. 
Therefore, the discrete eigenfunction corresponding to $\omega=0$, is given by  
\begin{equation}
	\label{zeromode}
	\npsi(x)=\nvarphi_{x}(x)= \sqrt{2\,U[\nvarphi(x)]}.
\end{equation} 
This result is a consequence of the translational invariance of Eq.\ (\ref{sg})
\cite{buratti:1968,scott:1969}. The study of the stability of pulses $\nphi_p(x)$ also leads to 
Eqs.\ \eqref{sg5}-\eqref{zeromode} changing $\nvarphi(x)$  to $\nphi_p(x)$. The relation $\nvarphi_{x}(x)= \sqrt{2\,U[\nvarphi(x)]}$ is usually known as the 
	Bogomolnyi equation \cite{bogomolnyi:1976,manton:2004}, although it appears  earlier in this context (see, for instance, Eq. (2.4) of Ref. \cite{goldstone:1975}). 

Since $\nvarphi(x)$ represents a kink, its derivative $\nvarphi_{x}(x)$ has no 
zero as it is shown, for instance, Eqs.\ \eqref{sgk} and \eqref{phi4k} for the sine-Gordon and $\varphi^4$ kinks. This means that $\omega^2=0$ is 
the lowest eigenvalue, and therefore all other eigenvalues are positive. 
For the pulses, however, $\npsi(x)=d\nphi_{p}/dx$ has at least one zero (see the third column of Table 
\ref{tabla2}) and, since there could be a negative eigenvalue, the positiveness of all the  eigenvalues cannot be guaranteed.

Given that $\omega^2$ is real, $\omega$ can be either an imaginary number or a
real number. The former case implies that $\Psi(x,t)$ in \eqref{ansatz} is unbounded when $t \to +\infty$ (the static solution  is unstable), while the latter case 
leads to a bounded function $\Psi(x,t)$ in $t$.  Is the boundedness of  $\Psi(x,t)$ a sufficient condition for a static kink or pulse to be stable? 
To answer this question, it is necessary to define what stability means. Here we generalize the concept of linear stability 
of nonlinear equations carried out in Ref. \cite{eckhaus:1965} for the Sturm-Liouville problem with only a denumerable set of eigenvalues.

To be precise, the static solution  of Eq. (\ref{sg}) 
is defined to be \textit{linearly stable} if all the solutions $\{\omega^2,\psi(x)\}$ of the associated
Sturm-Liouville problem (\ref{sg8}) which belong to $C^2_\infty(\mathbb{R})$ have real $\omega$. 
 From this  definition, it follows that the static solution is stable if all eigenvalues 
$\omega^2$ are non-negative \cite{goldstone:1975}, 
that is, $\omega$ is real, on the condition that $\psi(x)\in C^2_\infty(\mathbb{R})$.

\section{Solving the Sturm-Liouville problem} \label{sec3}

In the previous section, we  restricted ourselves to discussing the time-dependent 
part of the solution (\ref{ansatz}),  and showed that, for sine-Gordon and $\varphi^4$ potentials, 
it is bounded. In addition, it is convenient to establish  whether the eigenfunctions of the Sturm-Liouville problem 
(\ref{sg8}) form an orthogonal and complete set. 
It is worth mentioning the importance of the completeness condition since, in practical applications, the function 
$\Psi(x,t)$ in Eq.\ \eqref{sg5} is expanded in the set  $\{\omega^2,\psi(x)\}$ \cite{eckhaus:1965,dauxois:2006}.

To this end, we rearrange the terms of Eq.\ (\ref{sg8}) as (see Eq.\ (2.1) of Ref. \cite{takh:2008})
\begin{equation}
	\label{sl-sim}
-\frac{d^2\psi(x)}{dx^2}+V(x)\psi(x)=k^2 \psi(x),
\end{equation}
where $k^2=\omega^2-\omega_{ph}^2$, the potential $V(x)$ is given by 
\begin{equation}
	\label{potenciales}
	V(x)=U''[\nvarphi(x)]-\omega_{ph}^2, 
\end{equation}
and  
\begin{equation}
	\label{lim}
	\omega_{ph}^2=\lim_{x \to - \infty} U''[\nvarphi(x)] 
\end{equation}
is a non-negative constant since kinks and pulses 
depart from one minimum (at $x \to -\infty$) of the potential $U[\nvarphi(x)]$. 
The advantage of writing the Sturm-Liouville problem in the form \eqref{sl-sim} is that 
the potential $V(x)$ approaches zero when $x \to -\infty$.
It is precisely the asymptotic behavior of the P\"oschl-Teller potential \cite{epstein:1930,eck:1930} 
\begin{equation}
	\label{utildep}
	V(x)=-\frac{l\,(l+1)}{\cosh^2(x)}, \quad l\in\mathbb{N},
\end{equation}
which is straightforwardly obtained 
for the sine-Gordon kink ($l=1$), the $\varphi^4$ kink ($l=2$), and for the pulses corresponding to the cubic ($l=3$) and quartic ($l=2$) potentials, see Table \ref{tabla2}. 
For other nonlinear Klein-Gordon potentials, the function $V(x)$ is more complicated (see, for instance, 
$V(x)$ associated with the $\varphi^6$  and with the double sine-Gordon equations 
in Refs. \cite{lohe:1979,campbell:1986}, respectively). 
Moreover, the function  \eqref{utildep} belongs to the class of potentials that satisfy the condition
\begin{equation}\label{con-pot}
\int_{-\infty}^{+\infty}(1+|x|)\, |V(x)|dx<\infty,
\end{equation}
that is, $|x\,V(x)| \to 0$  when $x \to \pm \infty$. Equation \eqref{sl-sim} is
an instance of the one-dimensional Schr\"odinger equation and has been studied extensively (see  
Chapter 3, \S2 of Ref. \cite{takh:2008} and references therein for example).
Here we will explicitly solve Eq. \eqref{sl-sim} where $V(x)$ is given by Eq. \eqref{utildep}
in terms of the Jacobi polynomials, and we will show that our set of solutions are, in fact,  
an orthogonal and complete set in $L^2(\mathbb{R})$ (the square integrable functions on $\mathbb{R}$).

At the boundaries, the solution of Eq.\ (\ref{sl-sim}) behaves in the same way as $\exp(i k x)$. 
Without any loss of generality,  the solution of Eq. (\ref{sl-sim}) can be written as  
\begin{equation}
	\label{sg10}
	\psi(x)=e^{i k x}\,F(x).   
\end{equation}
By  substitution of Eq.\ \eqref{sg10} into  Eq. (\ref{sl-sim}), we obtain    
\begin{equation}
	\label{sg11}
	\frac{d^2 F(x)}{dx^2}+2 \,i\,k\,\frac{dF(x)}{dx}-V(x)\,F(x)=0. 
\end{equation}
 Consequent to the change of variable $s=\tanh(x)$, the domain of the function $F(s)$ reduces to $s \in (-1,1)$, and 
 Eq.\ (\ref{sg11}) with the potential (\ref{utildep}), for $l\in\mathbb{N}$, can be rewritten as 
\begin{equation}
	\label{sl-jac}
	(1-s^2)\,\frac{d^2 F(s)}{ds^2}-2 (s-i\,k)\,\frac{dF(s)}{ds}+l(l+1)\,F(s)=0,
\end{equation} 
which is the Jacobi differential equation, see Eq. (4.2.1) on page 60 in Ref. \cite{szego:1959} 
with $\alpha=-i\,k$ and $\beta=i\,k$. Equation \eqref{sl-jac} has two linearly independent solutions.  
Its only bounded solution is the Jacobi polynomial  $P_l^{(-i k,i k)}(s)$, 
see \S4.2 on page 60-62 in Ref. \cite{szego:1959}, 
where $l$  represents the degree of the polynomial.  For more details on Eq.\ \eqref{sl-jac}  
and on the Jacobi polynomials, the reader is referred to the books \cite{szego:1959,rusev:2005}, 
as well as the handbook \cite{olver:2010}. 

Hence, a solution of Eq.\ \eqref{sl-sim} is
\begin{equation}
	\label{eqsol}
	\psi(x)=e^{i k x}\,P_l^{(-i k,i k)}(\tanh x).   
\end{equation}
The parameter  $k$ can be either real or  imaginary (recall that 
$k^2$ is real). We consider these two cases separately.

\subsection{$k \in \mathbb{R}$ and the continuous spectrum}

By assuming, first, that  $k\in\mathbb{R}^{+}\backslash\{0\}$ in Eq.\ \eqref{sl-sim}, then the frequencies of the continuous spectrum  
 $\omega=\omega(k)=\sqrt{\omega_{ph}^2+k^2}$ are obtained. Two  bounded solutions of Eq.\ \eqref{sl-sim} are given by 
\begin{eqnarray}\label{sl-cI}
	\psi(x,k)&=&e^{ikx} P_l^{(-i k,ik)}(\tanh x),  
\end{eqnarray}
and its complex conjugate $\overline{\psi(x,k)}=\psi(x,-k)$, where $\overline{A}$ represents the complex conjugate of $A$. Direct calculations show that  $\psi_{x}(x,k)$ and $\psi_{x}(x,-k)$ are also bounded. 

 We now show that $\psi(x,k)$ and $\psi(x,-k)$  are two 
independent solutions of Eq. \eqref{sl-sim}. Indeed, by calculating the Wronskian
\begin{equation}\label{wro}
W[\psi(x,k),{\psi(x,-k)}]=
\det  \left|
\begin{array}{cc}
	\psi(x,k)     &  {\psi(x,-k)} \\ 
	\psi_{x}(x,k)    &  {\psi_{x}(x,-k)}
\end{array}
\right|,
\end{equation}
it is straightforward to show that 
\begin{eqnarray} \nonumber
W[\psi(s,k),{\psi(s,-k)}]=  -2ik   P_l^{(-i k,ik)}(s) P_l^{(i k,-ik)}(s) \\  \label{w1}+
(1-s^2)W[ P_l^{(-i k,ik)}(s), P_l^{(i k,-ik)}(s)],
\end{eqnarray}
where, for simplicity, the Wronskian is written in the variable $s=\tanh x$. 
Taking the limit $s \to  1$ ($x\to+\infty$) in Eq.\ \eqref{w1}, and using   
\begin{equation}
	P_l^{(\nu,-\nu)}(1)= \frac{(\nu+1)(\nu+2)\cdots(\nu+l)}{l!}, \label{jac+1}
\end{equation}
it is determined  that,  
\begin{equation}\label{val-wro}
\lim_{x\to+\infty} W[\psi(x,k),{\psi(x,-k)}]
={-2ik }{A_{l,k}^2}\neq0,
\end{equation}
since  $k \ne 0$, and 
\begin{equation}\label{A_k}
A_{l,k}^2:=\frac{1}{l!^2}\prod_{m=1}^l (k^2+m^2) 
\end{equation} 
is always a positive constant. Recall that, by Liouville's formula, %\cite[\S27.6]{arnold:2006}
	see \S27.6 in Ref. \cite{arnold:2006}, 
the Wronskian \eqref{wro} is independent of $x$, and, therefore, takes the value given by Eq. \eqref{val-wro} for all $x\in\mathbb{R}$.

It suffices to consider $k>0$ since $\psi(x,k)$ transforms into $\psi(x,-k)$ if $k\to-k$. 
From the well-known result from the Sturm-Liouville theory 
(see  \S15 in Ref. \cite{simmons:1991}), it follows that the general solution of 
Eq. \eqref{sl-sim}  is, therefore, a linear combination of $\psi(x,k)$ and $\psi(x,-k)$.

 For the specific value of $k=0$,  the  only bounded solution, $\psi(x,0)=P_l^{(0,0)}(\tanh x)$, is the Legendre polynomial of degree $l$ 
(see Theorem 4.2.2 on page 61 and the subsequent discussion in Ref. \cite{szego:1959}). 
This eigenfunction corresponds to the  lowest frequency of the continuous spectrum, $\omega(0)=\omega_{ph}$.

Therefore, the continuous spectrum $k \in [0,\infty)$ of Eq. \eqref{sl-sim} is characterized, up to constant 
factors, by the functions $\psi(x,\pm k)$,  $k \ge 0$, where
\begin{equation}\label{continuo}
\psi(x,k)=e^{ikx} P_l^{(-i k,ik)}(\tanh x), \quad  \omega(k)=\sqrt{\omega_{ph}^2+k^2}.
\end{equation} 

\subsection{ $k=i\,\kappa$, $\kappa\in\mathbb{R}^{+}\backslash\{0\}$, and the discrete spectrum}

Let us consider  the second case in which $k=i\,\kappa$ is a pure imaginary number, where  $\kappa\in\mathbb{R}^{+}\backslash\{0\}$. From Eq. \eqref{eqsol}, one solution  of the  Sturm-Liouville problem \eqref{sl-sim}  has the form 
\begin{equation}\label{sl-cII}
	\psi_\kappa(x)=e^{-\kappa x} P_l^{(\kappa,-\kappa)}(\tanh x), 
\end{equation}
where  $ P_l^{(\kappa,-\kappa)}(\tanh x)$ is a bounded function in $\mathbb{R}$. 
When $x\to+\infty$,  $\psi_\kappa(x)$ and $\frac{d}{dx}\psi_\kappa(x)$  go to zero. 
However, when $x\to-\infty$, the exponential function $e^{-\kappa x}$ goes to infinity. 
Therefore, $\psi_\kappa(x)$  is bounded if the following condition is satisfied
\begin{equation}
	\label{limA}
	\lim_{x \to -\infty}P_l^{(\kappa,-\kappa)}(\tanh x)=0.
\end{equation}
Using the value 
\begin{equation}
	P_l^{(\nu,-\nu)}(-1)=  \frac{(\nu-1)(\nu-2)\cdots(\nu-l)}{l!},\label{jac-1}
\end{equation}
it can be shown that Eq.\ (\ref{limA}) holds if and only if $\kappa=1,2,\ldots,l$. 
Given that $\kappa$ takes on $l$ discrete values, 
this case leads to the discrete spectrum. Hence, if the condition
\begin{equation}	\label{d1}
	\lim_{x\to -\infty} e^{-\kappa x} P_l^{(\kappa,-\kappa)}(\tanh x)=0,\quad \kappa=1,2,\ldots,l,
\end{equation}
holds, then $\psi_\kappa(x)$ is bounded.  

 It is convenient to render the change of variable $x\to-x$ in Eq. (\ref{d1}), and then to use the symmetry property
\begin{equation}
	P_l^{(\kappa,-\kappa)}(-s)=(-1)^l P_l^{(-\kappa,\kappa)}(s),
	\label{sim-j}
\end{equation}
in order to obtain the following equivalent condition  of Eq.\ \eqref{d1}  
\begin{equation}	\label{d1eq}
	\lim_{x\to + \infty} e^{\kappa x} P_l^{(-\kappa,\kappa)}(\tanh x)=0,\quad \kappa=1,2,\ldots,l.
\end{equation}
We prove the  condition \eqref{d1eq} in two steps. First, by the changing of variable $t=e^{2x}$ in 
the l.h.s. of Eq.\  \eqref{d1eq}, and second,  by using the explicit expression of the Jacobi polynomial, 
see  Eq. (4.22.2) on page 64 \cite{szego:1959}, for $\kappa=1,2,\ldots,l$     
\begin{equation}\label{pol-jac-N}
	P^{(-\kappa,\kappa)}_{l}\left(s\right)=  {l \choose \kappa}^{-1}  {l+\kappa \choose \kappa} \left(\frac{s-1}{2}\right)^\kappa
	P^{(\kappa,\kappa)}_{l-\kappa}\left(s\right),
\end{equation} 
we find
\begin{eqnarray*}
	&& \lim_{t\to+\infty}  t^{{\kappa}/{2}}  P_l^{(-\kappa,\kappa)}\left(\frac{t-1}{t+1}\right)= \\
	&& \lim_{t\to+\infty} {(-1)^\kappa} 
	{l \choose \kappa}^{-1}  {l+\kappa \choose \kappa} P^{(\kappa,\kappa)}_{l-\kappa}\left(\frac{t-1}{t+1}\right) \frac{ t^{{\kappa}/{2}}}{(1+t)^\kappa},
\end{eqnarray*}%
which is equal to zero, taking into account that $P^{(\kappa,\kappa)}_{l-\kappa}(1)={l \choose \kappa}$. 
Here,  ${l \choose \kappa}=\frac{l!}{\kappa! (l-\kappa)!}$ is the binomial coefficient. 
The same procedure can be employed to show that $\lim_{x\to-\infty} \frac{d\psi_\kappa}{dx}(x)=0$. 
Therefore, $\psi_\kappa(x)$ and its derivative are bounded provided that $\kappa=1,2,\ldots,l$. 
Even though the function $\psi_{-\kappa}(x)=e^{\kappa x}$ $P_l^{(-\kappa,\kappa)}(\tanh x)$ 
is also a bounded solution of Eq. (\ref{sl-sim}) when $\kappa=1,2,\ldots, l$,  
 we do not take it into account owing to the fact that $\psi_\kappa(x)$ and $\psi_{-\kappa}(x)$ are linearly 
dependent since the Wronskian satisfies $\lim_{x \to + \infty} W[\psi_\kappa(x),\psi_{-\kappa}(x)]=0$. 

Let us now show that  the second linearly independent solution of (\ref{sl-sim}), 
here denoted by $\chi(x)$, is disregarded because either it is unbounded  or  
its first derivative is unbounded. Since $\psi_{\kappa}(x)$ and $\chi(x)$ are two linearly independent solutions of (\ref{sl-sim}), its Wronskian
\begin{equation}
	W[\psi_{\kappa}(x),\chi(x)] =  
	\psi_{\kappa}(x) \frac{d\chi}{dx}(x) - \chi(x)\frac{d\psi_{\kappa}}{dx}(x)
\end{equation}
must be different from zero for all $x \in \mathbb{R}$. Under the hypothesis 
that $\chi(x)$ and $\frac{d}{dx}\chi(x)$ are bounded, and taking into account 
that $\lim_{x\to-\infty}\psi_\kappa(x)=\lim_{x\to-\infty}  \frac{d\psi_\kappa}{dx}(x)=0$, 
it follows that $\lim_{x\to-\infty} W[\psi_\kappa(x),\chi(x)] =0$, which is a contradiction. 
Therefore, the hypothesis on  $\chi(x)$ and $\frac{d}{dx}\chi(x)$ is false, and at 
least one of these functions must be unbounded. 

With the help of Eq. \eqref{pol-jac-N}, the discrete eigenfunctions \eqref{sl-cII} can be rewritten, 
for $\kappa=1, 2,  \ldots, l$, as 
\begin{equation}\label{discrete}
	\psi_\kappa(x)= N_\kappa e^{-\kappa x} (1+\tanh x)^\kappa
	P^{(\kappa,\kappa)}_{l-\kappa}\left(\tanh x\right),   
\end{equation}
where $\omega_{\kappa}=\sqrt{\omega_{ph}^2-\kappa^2}$, and $l$ represents the number of discrete modes.  
The normalizing constants,  $N_\kappa$, are determined such that the $L^2$-norm 
$\|\psi_\kappa\|^2:=\int_\mathbb{R} |\psi_\kappa|^2dx=1$. 

\subsection{Completeness of the set of orthogonal eigenfunctions for all $ l \in \mathbb{N}$}
  \label{sec-comple}

After the Sturm-Liouville problem is completely solved, it is necessary to study  
the orthogonality and completeness of its set of eigenfunctions \eqref{continuo} 
and \eqref{discrete}.
In fact, in  Ref. \cite{takh:2008} it is shown: (i) that the eigenvalue problem \eqref{sl-sim}, 
where $V(x)$ satisfies the condition \eqref{con-pot}, has a real spectrum (and hence $k^2\in\mathbb{R}$, 
which agrees with our ansatz \eqref{ansatz}); and (ii) that there is a complete set of eigenfunctions
in $L^2(\mathbb{R})$. Since the P\"oschl-Teller potential fulfills \eqref{con-pot}, we can use 
the results in  Ref. \cite{takh:2008} to construct this complete set. 

In fact, the functions 
\begin{equation}\label{continuo-equiv}
	\widehat\psi(x,k)=\frac{e^{ikx}}{\sqrt{2\pi}\,A_{l,k}} P_l^{(-i k,ik)}(\tanh x), \quad  \quad k\in\mathbb{R},
\end{equation} 
together with the set given in Eq. \eqref{discrete}, satisfy the following completeness relation 
valid for all $\Phi(x)\in L^2(\mathbb{R})$
\begin{eqnarray}\nonumber
	\Phi(x)= & 
	\displaystyle	\int_{\mathbb{R}} \left[ \left(\int_{\mathbb{R}} \overline{\hpsi(y,k)}\Phi(y)dy \right) \hpsi(x,k) \right]dk \\ & \label{exp-psi}
	\qquad \displaystyle + \sum_{\kappa=1}^l   \left(\int_{\mathbb{R}} \psi_\kappa(y)\Phi(y)dy \right)  \psi_\kappa(x),
\end{eqnarray}
or, equivalently,
\begin{equation}
	\label{rel-com-psi} 
	\int_{\mathbb{R}}  \overline{\hpsi(y,k)} \hpsi(x,k) dk + \sum_{\kappa=1}^l  \psi_\kappa(x) \psi_\kappa(y)=\delta (x-y).
\end{equation}
Furthermore, the following the orthogonality relations hold:
\begin{equation}
	\label{rel-ort-psi-b-equ}\int_{\mathbb{R}} \psi_\kappa(x) \psi_\nu(x)dx=\delta_{\kappa,\nu},\quad 
	\int_{\mathbb{R}} \psi_\kappa(x)  {\hpsi(x,k)}dx=0,   
\end{equation}
\begin{equation}
	\label{rel-ort-psi}% \frac{A_k}{2\pi}
	\int_{\mathbb{R}} \overline{\hpsi(x,k)} \hpsi(x,m)dx=\delta (k-m),  
\end{equation}
where $k,m\in\mathbb{R}$ and $\nu,\kappa=1,2\ldots,l$.  The detailed proof 
 can be found in Appendix \ref{app-close}.

\subsection{Some examples}

\subsubsection*{The sine-Gordon equation ($l=1$).}

Equation \eqref{sg} with potential \eqref{sgp} is known in the literature as the sine-Gordon equation. 
Equation \eqref{sgk} represents its 
static kink solution.  
In the study of the linear stability of this topological wave, it is necessary to solve Eq. (\ref{sl-sim}) 
with the potential (\ref{utildep}) with $l=1$ and $\omega_{ph}=1$. 
Setting $\kappa=1$ in Eq.\ (\ref{discrete}), and taking into account that $P_0^{(1,1)}(s)=1$, the 
only discrete mode reads
\begin{equation}\label{discrete1}
	\psi_1(x)=\frac{1}{\sqrt{2}\cosh(x)}, \qquad \omega_{1}=0, 
\end{equation}
and corresponds to the aforementioned zero mode \cite{rubinstein:1970}. 

The eigenfunctions associated  to the continuous spectrum are given by Eq.\ (\ref{continuo-equiv}) with $l=1$:
\begin{eqnarray}\label{sl-cIsg1}
	\psi(x,k)&= 
	\frac{e^{ikx}\,[\tanh(x)-i\,k]}{\sqrt{2\pi} \omega(k)} , \quad  \omega(k)=\sqrt{1+k^2}, 
\end{eqnarray} 
where the value $P_1^{(-\nu,\nu)}(s)=s-\nu$ has been employed. 

From the relations \eqref{rel-com-psi}-\eqref{rel-ort-psi-b-equ},  the orthogonality and completeness relations are deduced:
\begin{equation*}
	\label{rel-ort-psi-sg}
	\int_{\mathbb{R}} \overline{\psi(x,k)} \psi_1(x) dx=0,  \quad  	\int_{\mathbb{R}} \overline{\psi(x,k)} \psi(x,m)dx=\delta (k-m),  
\end{equation*}
and
\begin{equation*}
	\label{rel-com-psi-sg}
	\psi_1(x)\psi_1(y)+\int_{\mathbb{R}} \overline{\psi(x,k)} \psi(y,k) dk=\delta (x-y),  
\end{equation*}
respectively, where $k,m\in\mathbb{R}$.
These relations are mentioned in  Ref. \cite{rubinstein:1970}. %without any proof. 
The expansion of the approximated solution of the perturbed sine-Gordon equation in terms of this set of functions \cite{fogel:1977,dauxois:2006} is now well-justified. 

\subsubsection*{The $\varphi^4$ equation ($l=2$).}

The stability of the $\varphi^4$ kink \eqref{phi4k} is determined by solving  Eq. (\ref{sl-sim}) with the potential (\ref{utildep}) 
with $l=2$ and $\omega_{ph}=2$. Therefore, there are two discrete modes since $\kappa=1, 2$. By setting $\kappa=1$ in Eq.\ (\ref{discrete}), 
and $P_1^{(1,1)}(s)=2s$, the so-called internal mode 
\begin{eqnarray}\label{discrete3}
	\psi_1(x)= \sqrt{\frac{3}{2}}\frac{\tanh(x)}{\cosh(x)}, \qquad 
	\omega_{1}=\sqrt{3}, \qquad 
\end{eqnarray}
is obtained. This is an odd function  
with only one zero. The existence of an internal mode explains the inelastic interaction between a kink and antikink of the $\varphi^4$ equation \cite{aubry:1976,campbell:1983}, and therefore prevents the integrability of the system  \cite{bogdan:1990,kevrekidis:2001,charkina:2006}.     

In the same way, by setting $\kappa=2$ in Eq.\ (\ref{discrete}), the translational mode reads
\begin{equation}\label{discrete2}
	\psi_2(x)=\frac{\sqrt{3}}{2\cosh^2(x)}, \qquad 
	\omega_{2}=0.  
\end{equation}
The continuous spectrum is above $\omega_{ph}=2$, and can be obtained from (\ref{continuo-equiv})
\begin{equation}\label{continuo-p4}
\psi(x,k)=\frac{e^{ikx}\left[3\tanh^2(x)-3ik \tanh(x)-k^2-1\right]}{\sqrt{2\pi(k^2+1)}\omega(k)} ,
\end{equation}
where $\omega(k)=\sqrt{4+k^2}$, and where the value $P_2^{(-\nu,\nu)}(s)=\frac32\left(s^2-\nu s+{{\nu^2-1}\over{3}}\right)$, for the second-degree Jacobi polynomial,
 is used.  The set of eigenfunctions  given by Eqs. (\ref{discrete3}), (\ref{discrete2}), and (\ref{continuo-p4}) agrees with 
 that obtained in Refs. \cite{dashen:1974,goldstone:1975,barashenkov:2009}. Moreover, from 
\eqref{rel-com-psi}-\eqref{rel-ort-psi-b-equ},  the orthogonality  and the completeness relations are found, 
\begin{equation*}
	\label{rel-ort-psi4}
	\int_{\mathbb{R}} \psi_\kappa(x) \psi(x,k)dx=0,  \quad	\int_{\mathbb{R}} \overline{\psi(x,k)} \psi(x,m)dx=\delta (k-m),  
\end{equation*}
and 
\begin{equation*}
	\label{rel-com-psi4}
	\psi_1(x)\psi_1(y)+\psi_2(x)\psi_2(y)+\int_{\mathbb{R}} \overline{\psi(x,k)} \psi(y,k) dk=\delta (x-y),  
\end{equation*} 
respectively, where $\kappa=1,2$ and $k,m\in\mathbb{R}$ (see Ref.\ \cite{bishop:1980}).

It is worthwhile to remark that not all the potentials of the form (\ref{utildep}) lead to linearly stable solutions of the nonlinear Klein-Gordon equation. 
Indeed, the pulses of the  cubic and quartic potentials are unstable. For instance, the stability of the former pulse is related with 
the P\"oschl-Teller potential with $l=3$. Since its lowest frequency of the continuous spectrum is 
$\omega_{ph}=2 < l$, the lowest eigenvalue 	$\omega^2_3=\omega_{ph}^2-l^2=-5$ is less than cero (see Table \ref{tabla2}). 
This analysis shows that, although all eigenfunctions and  their first derivatives are bounded 
(necessary condition for stability), the pulse is unbounded when $t \to +\infty$,  owing to a negative eigenvalue $\omega^2<0$. 	 

At this point, it is interesting to pose the following question:  further to the sine-Gordon and $\varphi^4$ kinks, and the  cubic and quartic pulses, are there other nonlinear Klein-Gordon kinks or pulses, whose stability is associated with 
 the Sturm-Liouville problem \eqref{sl-sim} with the P\"oschl-Teller potential (\ref{utildep})?

\section{From the P\"oschl-Teller potential to nonlinear Klein-Gordon potentials} \label{sec4} 
 
In order to answer the above question, let us consider two possibilities related with the lowest eigenvalue   $\omega_l^2$: (i) when
 $\omega_l=0$, that is, it agrees with the zero frequency; and (ii) if  $\omega_l^2 < 0$.
 The former case is related with the kinks, whereas the latter is related  with  the  pulses.
 
 Straightforward calculations show that 
 one of the solutions of the Sturm-Liouville problem \eqref{sl-sim} with the P\"oschl-Teller potential \eqref{utildep} reads \cite{morse:1953} 
 \begin{equation}
 	\label{psi1}
 	\psi_0^{(l)}(x)=\frac{A}{\cosh^l(x)}, \qquad \omega_l=\sqrt{\omega_{ph}^2-l^2}, 
 \end{equation} 
 where $A$ is a constant. The function $\psi_0^{(l)}(x)$ has no zeros, 
 therefore $\omega_l^2$ is the value   associated to the lowest eigenvalue.
 
 \subsection{The lowest eigenvalue corresponds to  the zero mode}  
 
 This case has been partially analyzed in Ref. \cite{christ:1975} for the values $l=1$ and $l=2$.
By setting $\omega_l=0$ in Eq.\ (\ref{psi1}),  it is implied that $\omega_{ph}=l$. 
From Eq. \eqref{zeromode}, and taking Eq. \eqref{sgbc2} into account, 
it follows that 
\begin{equation} \label{eqdd}
 \psi_l(x)=\frac{d}{dx}\varphi^{st}(x)=\frac{A_{l}}{\cosh^l(x)},
\end{equation} 
where 
\begin{equation*}
A_{l}=\frac{Q\,\Gamma(\frac{l+1}{2})}{\sqrt{\pi} \Gamma(\frac{l}{2})}.
\end{equation*}
By integrating Eq. \eqref{eqdd}, one obtains the solution of the nonlinear Klein-Gordon Eq.\  \eqref{sg} (without even knowing the potential).  This fact was already noticed in Ref. \cite{magyari:1985}, where, by using the Bogomolnyi equation (energy integral), the eigenvalue problem \eqref{sg8} was transformed into the second-order differential equation in the variable $\varphi$, and  was then solved. 
 
 Denoting this solution as $\varphi^{st}_{l}(x):=\varphi^{st}(x)$, it reads 
\begin{equation} \label{kinkl}
\varphi_{l}^{st}(x)=\varphi_{l-2}^{st}(x)+\frac{A_{l}}{l-1}   \frac{\sinh(x)}{\cosh^{l-1}(x)}, \qquad l \ge 2,
\end{equation} 
where $\varphi_{0}^{st}(x)=0$, $\varphi_{1}^{st}(x)=4\,\arctan[\exp(x)]$. This recurrence relation 
enables all the kink solutions to be systematically obtained. By rescaling the spatial variable with $l$, all the kink solutions for $l=1, \dots, 6$ of Table II of Ref. \cite{trullinger:1987} are recovered. Contrary to 
the kinks represented by Eq. \eqref{kinkl}, the width of the kinks of Table II of Ref. \cite{trullinger:1987} increases as odd (even) values of $l$ increase.

 In this circumstance, the family of solutions $\varphi_l^{st}(x)$ is linearly stable, and its energy \eqref{H-E} 
\begin{equation} \label{energy}
	H_{l}=\frac{Q^2}{\sqrt{\pi}}\frac{\Gamma(l)\,\Gamma^2(\frac{l+1}{2})}{\Gamma(l+\frac{1}{2})\,\Gamma^2(\frac{l}{2})},
\end{equation}
represents the Bogomolnyi bound \cite{manton:2004}.
Furthermore, from (\ref{zeromode}) and (\ref{psi1}), we obtain  the potential 
 \begin{equation}
 	\label{eqUphi0}
 	U[\varphi^{st}(x)]=\frac{A^2_{l}}{2\,\cosh^{2l}(x)}. 
 \end{equation} 
In the following, for the sake of brevity,  henceforth $U$ denotes the potential function $U[\varphi^{st}(x)]$, that is, $U:=U[\varphi^{st}(x)]$.
By inserting Eq.\ (\ref{utildep}) into (\ref{potenciales}),  and by using (\ref{eqUphi0}), it 
is straightforward to see that the nonlinear Klein-Gordon potential satisfies the following second-order differential equation
 \begin{equation}
 	\label{eqd}
 	U''+l\,(l+1)\,\alpha^2 U^{1/l}=l^2,
 \end{equation}
 where $\alpha^2= (2/A^2_{l})^{(1/l)}$. 
 This equation has been solved by using 
 	the Student's $t$-distribution in Ref. \cite{trullinger:1987}, where the cases of even and odd values of $l$ were analyzed separately. Here, we provide a more direct way to solve this equation. Indeed, we write the solution in terms of the Gauss hypergeometric function
 	${_2}F_1$ which is more familiar to a wider audience. In fact, the
 	Student's $t$-distribution
 	is usually expressed in terms of the hypergeometric function , see e.g., Ref. \cite{amos:1964}.
 
By multiplying this equation by $U'(\nvarphi)$ and integrating  the following first-order separable differential equation is obtained
 \begin{equation}
 	\label{eqd1}
 	U'^{2}=2\,l^2\,U\left(1- \alpha^2\,U^{1/l}\right),
 \end{equation} 
whose solution, by quadrature, is
\begin{equation}
\label{eqd1a}   
\int_0^U \frac{dt}{t^{1/2}(1-\alpha^2 t^{1/l})^{1/2}}=\pm \sqrt{2}l[C_{\pm}-\varphi^{st}(x)],
\end{equation} 
whereby $C_{\pm}$ is an integration constant. Making the change of variable $t=U\xi^l$ in the integral, and using the
Eq. (15.6.1) on page 388 of Ref. \cite{olver:2010} it follows that 
 \begin{eqnarray}
 	\label{eqd2}
 %	\sqrt{U\left(1-\alpha U^{1/l}\right)} {}_2F_{1}\left(1,\frac{l+1}{2};\frac{l}{2}+1;\alpha U^{1/l}\right)&=&\frac{1}{\sqrt{2}}\,l\,(\varphi_0-C),\\
 	\sqrt{2U}  {}_2F_{1}\left(\frac{l}{2},\frac{1}{2};\frac{l}{2}+1; \alpha^2\,U^{1/l}\right)&=&\pm l[C_{\pm}\!-\varphi^{st}(x)],
 \end{eqnarray} 
where ${}_2F_{1}$ denotes the hypergeometric function, see Chapter 15 in Ref. \cite{olver:2010}. The constants  $C_{\pm}$ can be chosen arbitrarily, and 
they set the value $\lim_{x \to \pm\infty} \varphi^{st}(x)=C_{\pm}$.  Notice that 
$C_{+}-C_{-}=Q$. 
This equation has two branches: one for the positive sign and the other for the negative sign. From the former, we obtain the 
part of the kink that  extends from the maximum of the potential to the second minimum ($U'<0$). From the latter, 
we calculate the part of the kink that  lies between the first minimum of the potential and the  maximum, that is $U'>0$.   

Equation \eqref{eqd2} defines the potential, at least implicitly, for all $l\in\mathbb{N}$. 
With the help of the following recurrence relation
\begin{eqnarray}\nonumber
&&	{_2}F_{1}\left(\frac{l}{2},\frac{1}{2};\frac{l}{2}+1; \alpha^2\,U^{1/l}\right)= \frac{l}{l-1}\frac{1}{\alpha^2 U^{1/l}}\times \\
 && \left[ {_2}F_{1}\left(\frac{l}{2}-1,\frac{1}{2};\frac{l}{2}; \alpha^2\,U^{1/l}\right)-\sqrt{1-\alpha^2 U^{1/l}}\right]\label{rr-2f1}
\end{eqnarray}
satisfied for $l \ge 2$, 
  Eq. \eqref{eqd2} can be solved for different values of $l$. This relation is obtained from the contiguous 
relation given by Eq. (15.5.16) on page 388 in Ref. \cite{olver:2010},
where $a=l/2$, $b=1/2$, $c=l/2$, $z=\alpha^2 U^{1/l}$, 
and by using the identity ${_2}F_1(a,1/2;a;z)=(1-z)^{-1/2}$ (see Eq. (15.4.6) in Ref.\ \cite{olver:2010}).

%\subsubsection{Case $l=1$}

\begin{figure}[h]
 	\begin{tabular}{cc}
 		\includegraphics[width=0.5\linewidth]{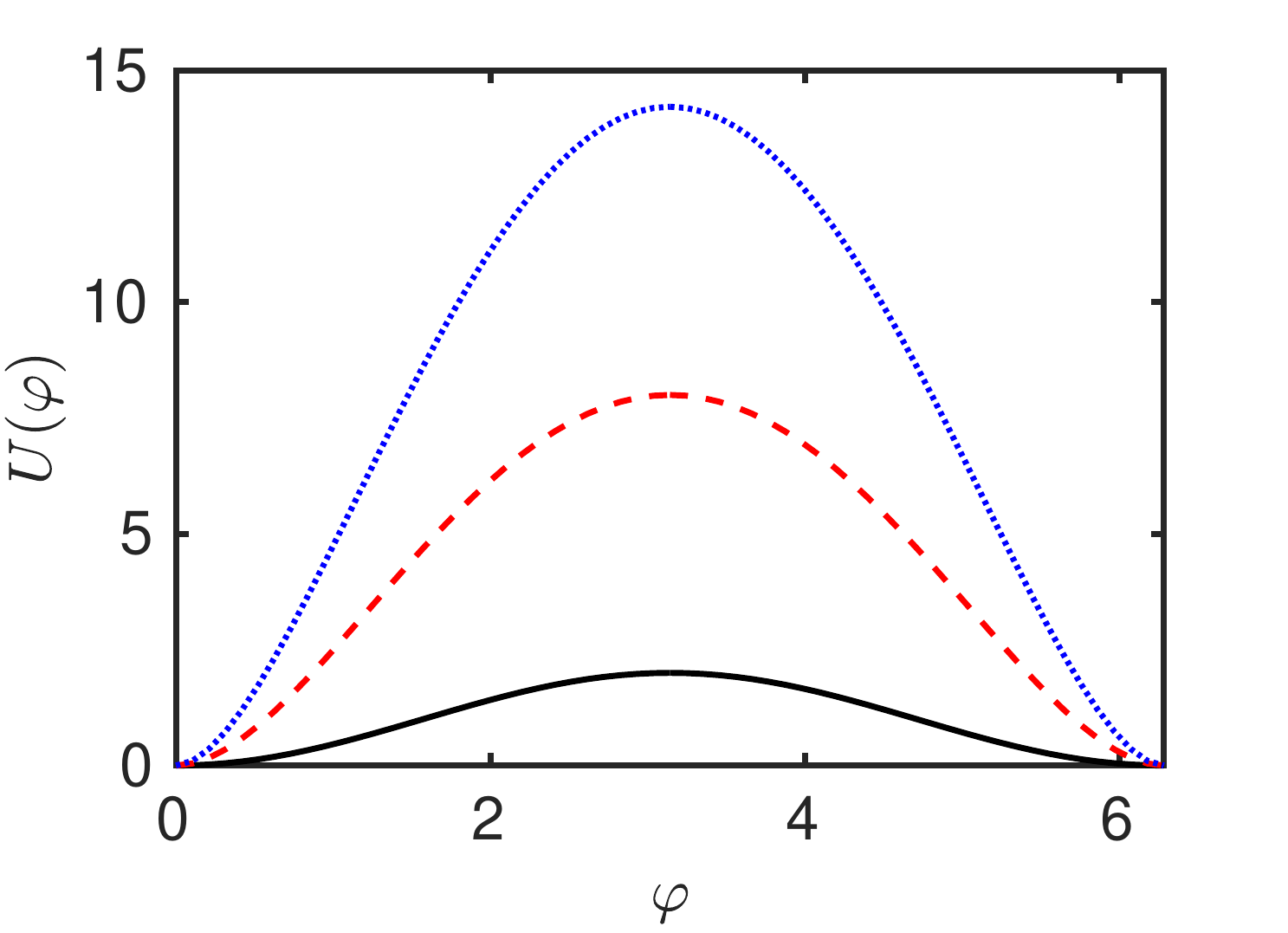} &
 		\includegraphics[width=0.5\linewidth]{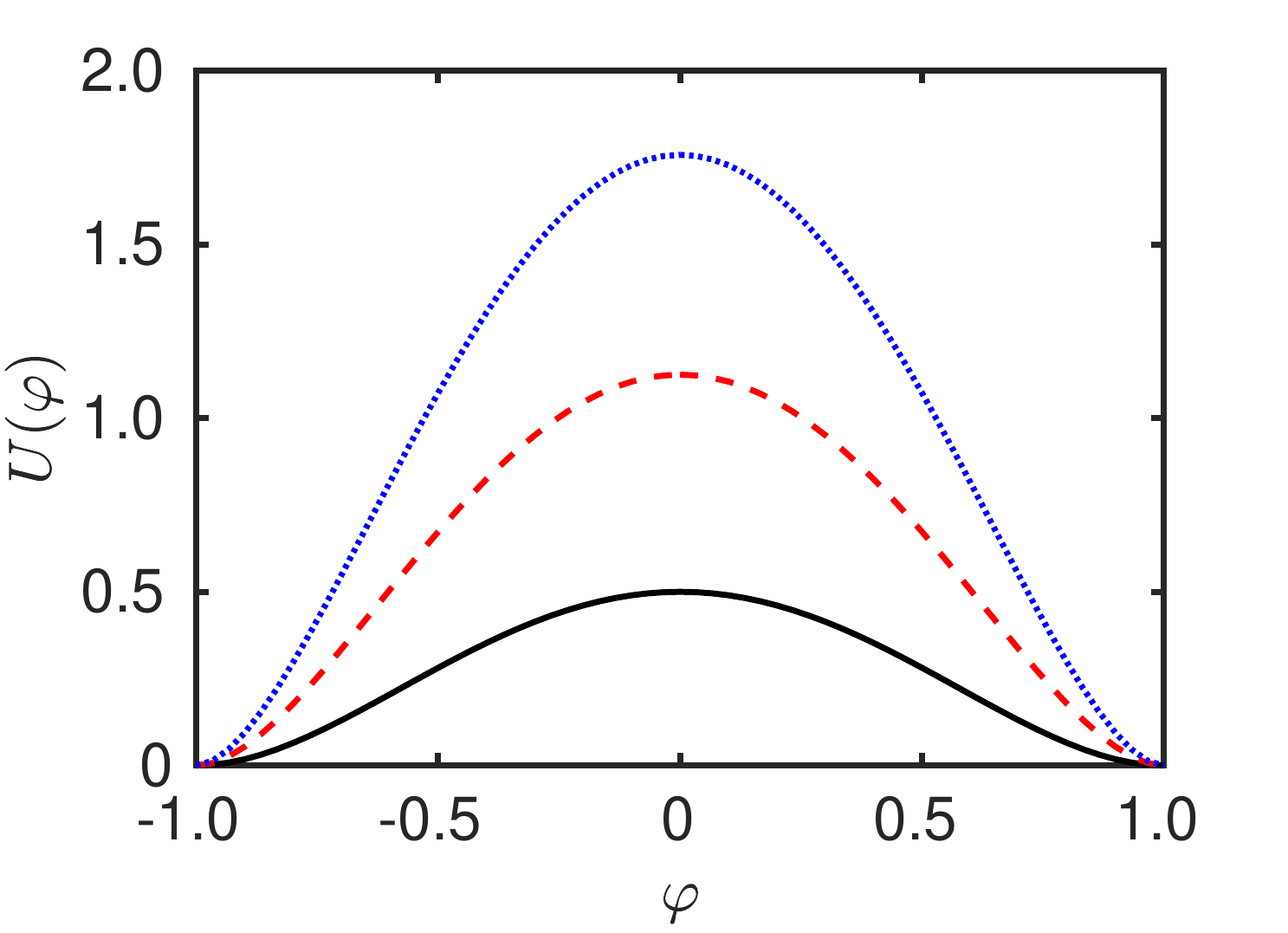}
 	\end{tabular}
 	\caption{ 
 		The nonlinear Klein-Gordon potentials are shown. Left-hand panel: $l=1$ (solid black line), $l=3$ 
 		(numerical results, dashed red line), and $l=5$ (numerical results, dotted blue line).
 		Right-hand panel: $l=2$ (solid black line), $l=4$ 
 		(dashed red line), and $l=6$ (numerical results, dotted blue line).}	
 	\label{fig2}
 \end{figure}

By setting $l=1$ in  Eq.\ \eqref{eqd2}  and using the identity (see Eq. (15.4.4) on page 386 in Ref. \cite{olver:2010}),  
\begin{equation} 
	\label{2f1-sv}{_2}F_{1}\left(\frac{1}{2},\frac{1}{2};\frac{3}{2}; z \right)=\frac{\arcsin(\sqrt{z})}{\sqrt{z}},\quad
\end{equation}
it follows that
\begin{eqnarray}
\nonumber
&&U[\varphi^{st}(x)]=\frac{1}{\alpha^2} \, \sin^2\left(\frac{\alpha (\varphi^{st}(x)-C_{\pm})}{\sqrt{2}}\right).
\end{eqnarray} 
By assuming $C_{-}=0$ and the topological charge $Q=2\pi$, then $\alpha=1/\sqrt{2}$, and  the sine-Gordon potential \eqref{sgp} is obtained, see Fig.\ \ref{fig2}.  
%\subsection{Case $l=2$}

Let us consider $l=2$. In this case the function $ {_2}F_1$ in Eq.\ \eqref{eqd2} reads
${_2}F_{1}\left({1},{1}/{2};{2}; z\right)$, $z=\alpha^2\sqrt{U}$. Using Eq. \eqref{rr-2f1} and taking into account that 
${_2}F_{1}\left(0,b;c; z\right)=1$, it follows that 
\begin{equation} 
\label{2f1-sv2}
{_2}F_{1}\left(1,\frac{1}{2};{2}; z \right)=\frac{2}{z}\left(1-\sqrt{1-z}\right).
\end{equation}
Therefore, Eq.\ \eqref{eqd2} gives
\begin{eqnarray}
	\nonumber
	%	\sqrt{U\left(1-\alpha U^{1/l}\right)} {}_2F_{1}\left(1,\frac{l+1}{2};\frac{l}{2}+1;\alpha U^{1/l}\right)&=&\frac{1}{\sqrt{2}}\,l\,(\varphi_0-C),\\
	&&   \frac{\sqrt{2 }}{\alpha^2}\,\left(1-\sqrt{1-\alpha^2 \sqrt{U}}\right)=\pm [C_{\pm}-\varphi^{st}(x)]. 
\end{eqnarray} 
Assuming $C_{-}=-1$, and $Q=2$, it follows that $\alpha^2=\sqrt{2}$ and $C_{+}=1$,  and we recover the $\varphi^4$ potential \eqref{phi4p}, see Fig.\ \ref{fig2}. 
%\begin{eqnarray}
%	U(\varphi_0)=\frac{(1-\varphi_0^2)^2}{2}, 
%\end{eqnarray} 
 
%\subsection{Case $l=3$}

\begin{figure}[h]
 	\begin{tabular}{cc}
		\includegraphics[width=0.5\linewidth]{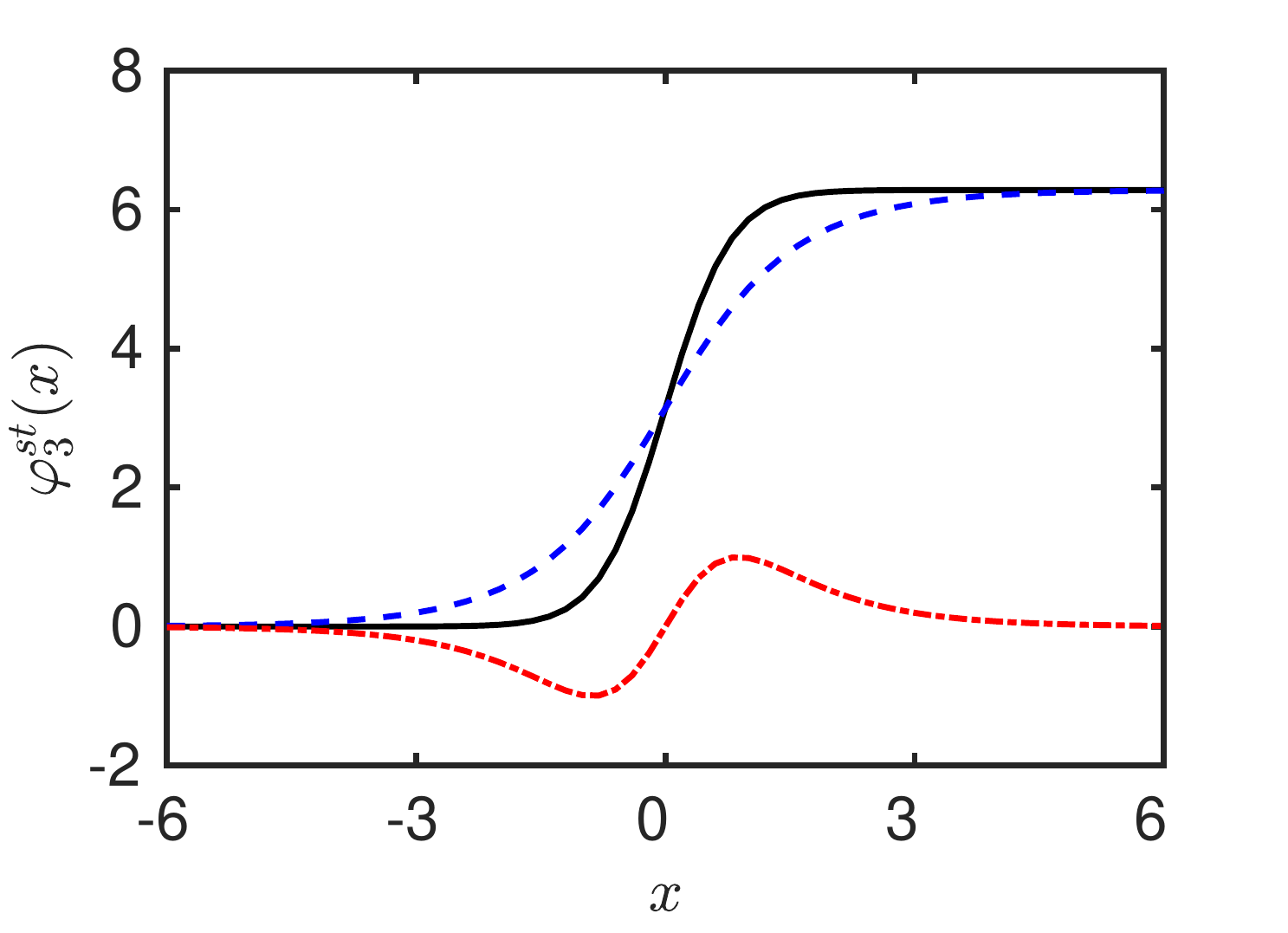} &
 		\includegraphics[width=0.5\linewidth]{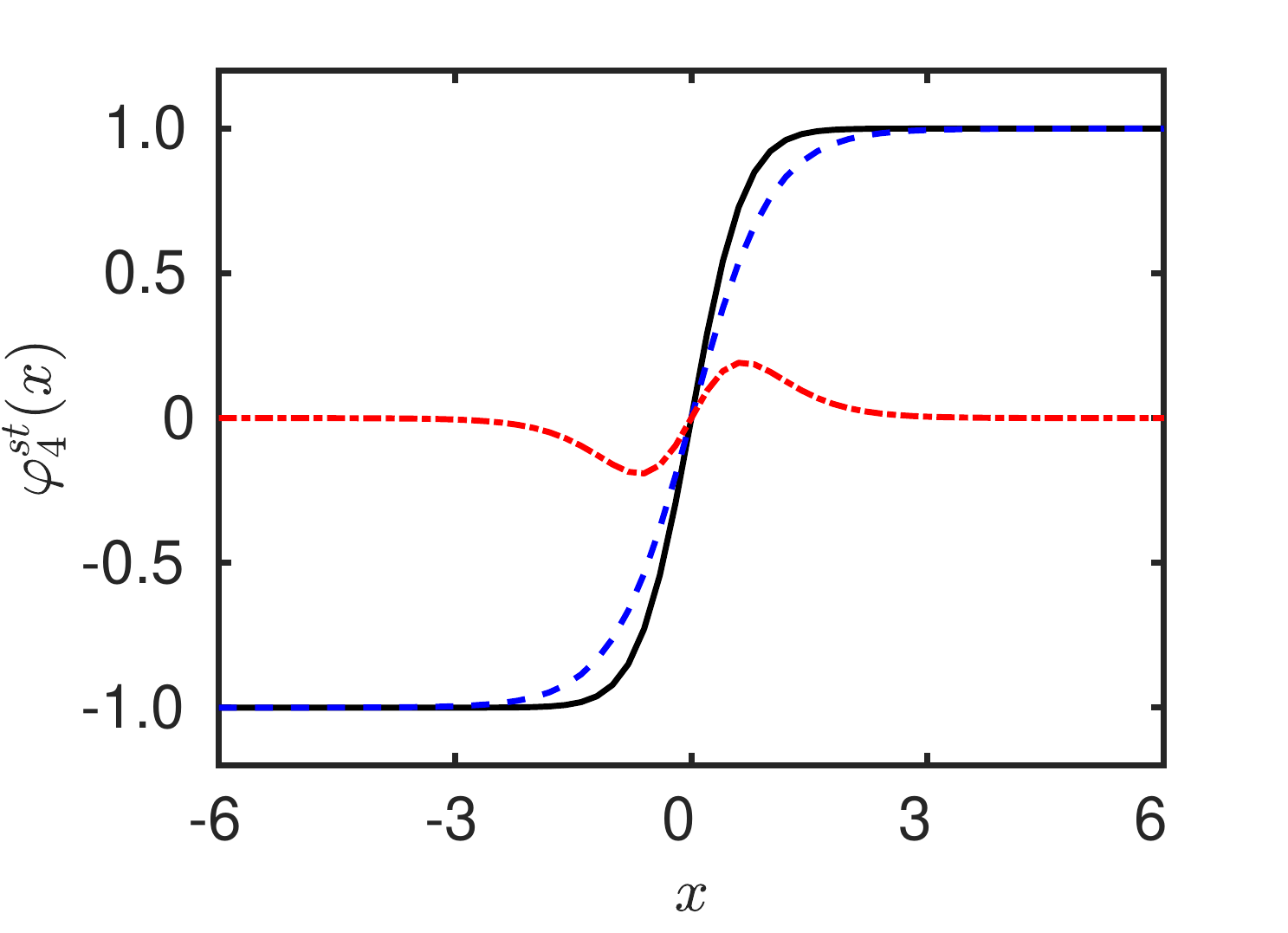}
 	\end{tabular}
 	\caption{Left-hand and right-hand panels show the kinks (solid black line), represented by Eq.\ \eqref{kinkl} for $l=3$ and $l=4$, respectively. This solution is a linear superposition of a kink 
 		(dashed blue line) and a localized function (dot-dashed red line).}
 	\label{fig3}
 \end{figure}

Setting $l=3$ in Eq.\ \eqref{eqd2}, and using Eqs.\  \eqref{rr-2f1}-\eqref{2f1-sv}, 
we obtain
\begin{eqnarray}
	\label{eql3}
	%	\sqrt{U\left(1-\alpha U^{1/l}\right)} {}_2F_{1}\left(1,\frac{l+1}{2};\frac{l}{2}+1;\alpha U^{1/l}\right)&=&\frac{1}{\sqrt{2}}\,l\,(\varphi_0-C),\\
%	&&\sqrt{U/2} \,\left(\frac{\arcsin(\alpha U^{1/6})}{\alpha^3 \sqrt{U}}-
%	\frac{\sqrt{1-\alpha^2 U^{1/3}}}{\alpha^2 U^{1/3}}\right) =\pm\,(C_{\pm}-\varphi^{st}(x)),  \\
\arcsin\left(\alpha U^{1/6}\right)&-&\alpha U^{1/6}
\sqrt{1-\alpha^2 U^{1/3}} =  \nonumber \\
&& \pm\sqrt{2}\,\alpha^3[C_{\pm}-\varphi^{st}(x)], 
\end{eqnarray} 
which has no explicit solution, see Fig.\ \ref{fig2}.    
%Here is assumed $C=-1$ and the topological charge $Q=2$, which implies $\alpha^2=\sqrt{2}$.   
 However, from Eq.\ \eqref{kinkl}, the kink solution (see Fig.\ \ref{fig3}) has the form  
\begin{eqnarray}
\label{kink3}
\varphi^{st}(x)=4\,\arctan[\exp(x)]+2\,\frac{\tanh(x)}{\cosh(x)},
\end{eqnarray}
where we set $Q=2\,\pi$.   
Moreover, we assume $C_{-}=0$, which implies $C_{+}=2\pi$, %$A=4$, 
and $\alpha=1/\sqrt{2}$. Interestingly, this kink (solid black line of Fig.\ 
\ref{fig3}) is a linear superposition of the sine-Gordon kink (dashed blue line of Fig.\ 
\ref{fig3}) and an odd localized function  in space (dot-dashed red line of Fig.\ \ref{fig3}). 
Since $l=3$, the lower phonon frequency is equal $\omega_{ph}=l=3$, and the $3$ discrete 
modes have  frequencies $\omega_{\kappa}=\sqrt{\omega_{ph}^2-\kappa^2}$, that is 
$\omega_{1}=2\,\sqrt{2}$, $\omega_{2}=\sqrt{5}$, and $\omega_{3}=0$. 
According to the results of the previous section, this kink is stable. 
 
Finally, let us consider the case of $l=4$.  We set $Q=2$, and $C_{-}=-1$, which imply %$A=3/2$, 
$\alpha=2^{3/8}/3^{1/4}$, and $C_{+}=1$. 
Equation \eqref{eqd2}, using Eqs.\  \eqref{rr-2f1} and \eqref{2f1-sv2}, 
%is obtained
%\begin{eqnarray}
%	\label{eql4}
%	&&\sqrt{2} \,\left(2-3\,(1-\alpha^2 U^{1/4})^{1/2}+(1-\alpha^2 U^{1/4})^{3/2}
%	\right) =\pm\,{3 \alpha^4 } (C_{\pm}-\varphi^{st}(x)). 
%\end{eqnarray}  
%In addition, specifying the values of $Q=2$, $C_{-}=-1$, which imply %$A=3/2$, 
%$\alpha=2^{3/8}/3^{1/4}$, and $C_{+}=1$. The Eq.\ \eqref{eql4} 
can be solved explicitly, and the potential, for $|\varphi^{st}(x)| \le 1$, has the form (see Fig.\ \ref{fig2})
\begin{eqnarray}\label{eql4}
U[\varphi^{st}(x)]=\frac{9}{8}\left[1-2 \cos\left(\frac{2}{3} \arcsin \varphi^{st}(x) \right)\right]^4.
\end{eqnarray} 
%This potential is represented in the right-hand panel of Fig.\ \ref{fig4}.
 From Eq. \eqref{kinkl}, its kink solution  
\begin{eqnarray}
	\label{kink4}
	\varphi^{st}(x)=\tanh(x)+\frac{1}{2}\,\frac{\tanh(x)}{\cosh^2(x)}
\end{eqnarray}  
is a linear superposition of two functions, the first one is the $\varphi^4$ kink, and the second one is a localized function (see right-hand panel of Fig.\ \ref{fig3}). 
 Since $l=4$, the lower phonon frequency  equals $\omega_{ph}=l=4$, and the $4$ discrete modes are related with the frequencies $\omega_{\kappa}=\sqrt{\omega_{ph}^2-\kappa^2}$, that is, $\omega_{1}=\sqrt{15}$, $\omega_{2}=\sqrt{12}$, $\omega_{3}=\sqrt{7}$, and $\omega_{4}=0$. Moreover, according to the results of the previous section, this kink is also stable. 

	Notice that the formula \eqref{eqd2} defines the Klein-Gordon potentials $U(\varphi)$, which can be 
explicitly expressed  for a few particular cases. From \eqref{eqd2} and \eqref{rr-2f1}, it can be shown that, for all odd values $l \ge 3$, it is impossible to find 
an explicit expression for $U$ since the functions $\arcsin z$ and $\sqrt{1-z}$, where $z=\alpha^2 U^{1/l}$, appear in different terms of the same equation.		
A similar situation happens when $l$ is an even number greater than $4$, since, in this 
case, the explicit solution is involved with the roots of a polynomial in $U$ of degree equal 
to or greater than $5$, which are, in general, impossible to
obtain analytically. %, so the only four explicit potentials are the previously obtained ones.
Therefore, for $l \ge 6$, the stable kink solution is represented by Eq.\ \eqref{kinkl}, whereas its corresponding Klein-Gordon potential can be numerically obtained by solving Eq.\ \eqref{eqd2}, and specifying  the topological charge $Q$ and the constant $C_{-}$.  

As a final remark it is important to point out the fact that Eq.\ \eqref{eqd1} was obtained in Ref.\ \cite{boya:1990} by differentiation of the positive branch of Eq. \eqref{eqd1a}. Since the authors analyzed  the first correction to the masses of a family of nonlinear Klein-Gordon kinks, rather than provide the solution of the differential equation for the potential, they calculated the kink's mass \eqref{H-E} by using the Bogomolnyi equation. 
	They  
	obtained $M_{l}=24^{l-1} \Gamma^2(l)/\Gamma(2\,l)$, which differs from the expression \eqref{energy} due to a different choice of the normalization constant $A_l$ in Eq.\ \eqref{eqdd}.  
	Here, $A_l$ is given by Eq.\ \eqref{eqdd} such that $Q=2$  for even values of $l$, whereas in Ref.\ \cite{boya:1990}, $A_l$ equals $1$ for all values of $l$.

\subsection{The lowest eigenvalue is negative} 

The second case deals with negative values of $\omega_l^2$, thus the solitary wave is linearly unstable.
Let us assume that the static pulse solution of  the nonlinear Klein-Gordon has the form
\begin{equation}
	\label{static}
	\phi^{st}(x)=\frac{1}{\cosh^n(x)},
\end{equation}
where the parameter $n$  is determined \textit{a posteriori}. Using the relationship \eqref{zeromode}, this condition implies that 
\begin{equation}  
	\label{c1}
	U(\phi)=%\frac{n^2}{2} \frac{1}{\cosh^{2n}(x)}\left(1-\frac{1}{\cosh^{2}(x)} \right)=
	\frac{n^2}{2} \phi^2 \left(1-\phi^{2/n} \right).
\end{equation}
The envelope part of the NLS soliton with arbitrary power-law nonlinearity $|\Psi(x)|^{2/n}$ is  represented by Eq. \eqref{static} since it satisfies the nonlinear Klein-Gordon equation \eqref{sg}, where the potential is given by Eq. \eqref{c1} \cite{sulem:1999}. However, the stability of the NLS soliton is determined by a more complex eigenvalue problem than that represented by Eq. \eqref{sl-sim}, see Chapter 4 of Ref. \cite{sulem:1999}. 
The investigation of the stability of the solution \eqref{static} leads to the Sturm-Lioville problem \eqref{sl-sim}, where $U''(\phi)=n^2-(n+1)(n+2)\phi^{2/n}$. By comparing this expression with Eqs.\ \eqref{potenciales}-\eqref{utildep}, we obtain $n=l-1$, and $\omega_{ph}^2=(l-1)^2$  
 ($l \ge 2$). From the above analysis, the discrete frequencies are represented by $\omega_{\kappa}^2=\omega_{ph}^2-\kappa^2$, where $\kappa=1, 2, \cdots, l$. Clearly, the frequency 
 $\omega_{l}^2=(l-1)^2-l^2<0$ and all pulses  \eqref{static} are unstable. 
 The question therefore arises as to whether there is any way to stabilize the pulses.

\section{Control of stability} \label{sec5}

The purpose of this section is to obtain stable pulses, associated to the nonlinear Klein-Gordon potential,   with the help of an inhomogeneous force $f(x)$.  
A similar procedure has been successfully  considered in
Refs. \cite{gonzalez:1998,gonzalez:2017} to control the existence of internal modes associated to topological solitons in the perturbed $\varphi^4$-potential and in the inhomogeneous sine-Gordon equation. 
For instance, let's consider the following nonlinear Klein-Gordon equation with an external force
\begin{equation} \label{eqq1}
	\phi_{tt}-\phi_{xx}+ \frac{d U}{d \phi}=f(x),
\end{equation}
where $U$ is the $\phi^3$ potential given by Eq.\ \eqref{c1} by setting $n=2$. 
The unstable static pulse solution, when $f(x)=0$, has the form $\phi^{st}(x)=1/\cosh^2(x)$.  Straightforward calculations show that the pulse 
\begin{equation} \label{solu}
\phi(x)=a\,\phi^{st}(b\,x), 
\end{equation}
with positive constants $a$ and $b$, is  the solution of Eq. (\ref{eqq1}) whenever 
\begin{equation} \label{eqqF}
f(x)=\frac{2a}{\cosh^4(bx)}\left[1-3\,a+2b^2 +(1-b^2)\cosh(2bx) \right].
\end{equation}
In order to study the stability of the pulse (\ref{solu}), 
the methodology of Section \ref{sec2} is applied. Hence, Eq.\    
(\ref{eqq1}) is linearized around the pulse, that is, the expansion (\ref{sg5}) 
is inserted in Eq.\ (\ref{eqq1}). %, and since   $\epsilon \ll 1$ is assumed, 
The function  $\Psi(x,t)$ satisfies Eq.\ (\ref{sg6}). Finally, by  
assuming the ansatz (\ref{ansatz}),  the function 
$\psi(x)$ satisfies the Sturm-Liouville problem	(\ref{sg8}). 

By inserting  the pulse (\ref{solu}) into the second derivative of the potential, it has the form 
\begin{equation}
	\label{segder}
	U''[\phi(x)]=4-\frac{12\,a}{\cosh^2(b\,x)}.
\end{equation}
By assuming  the change of variable $X=b\,x$, the Sturm-Liouville problem reads
%Considering (\ref{segder}), after the change of variable $X=B\,x$, the Sturm-Liouville equation reads
\begin{equation}
	\label{sleq}
	\psi_{XX}+\left[\frac{\omega^2}{b^2}-\frac{4}{b^2}+\frac{12\,a}{b^2\,\cosh^2(X)}\right]
	\psi=0.
\end{equation}
For certain values of $a$ and $b$, the $\sech^2(x)$ potential becomes the P\"oschl-Teller potential, that is,  
\begin{equation} \label{rel}
	\frac{12\,a}{b^2}=l\,(l+1). 
\end{equation}
There are two different ways to stabilize the pulse. To start with, the value of $b$ is fixed, for instance  as $b=1$. This implies that the lowest phonon frequency is $\omega_{ph}=2$. The discrete set of frequencies is given by 
$$
\omega^2=\omega_{ph}^2-\kappa^2, 
$$  
where $\kappa=1, 2, \cdots, l$. Demanding stability, all values of $\omega^2$ should be non-negative. This implies that $\min(\omega^2)=4-l^2 \ge 0$, i.e. either $l=1$ ($a=1/6$) or $l=2$ ($a =1/2$). In particular, 
the discrete mode for the case $l=1$ reads
\begin{equation}\label{disa}
	\psi_1(x)=e^{-x} P_1^{(1,-1)}(\tanh x)=\frac{1}{\cosh(x)}, \qquad \omega_{1}^2=3,
\end{equation}
while the continuous spectrum is represented by 
\begin{eqnarray}\label{sl-a}
	\psi_k(x)&=& %e^{ikx} P_1^{(-i k,ik)}(\tanh x)=
	e^{ikx} \,[\tanh(x)-i\,k], \qquad  \omega_{k}^2=4+k^2.  
\end{eqnarray} 
For the case $l=2$ ($a=1/2$), the solution of the Sturm-Liouville problem is represented by the expressions 
(\ref{discrete3}), (\ref{discrete2}), and (\ref{continuo-p4}). Hence, the two pulses considered are stable. 

The second method to stabilize the pulse is to set $a$, for instance $a=1/3$, and to change $b$ 
 in accordance with 
Eq. (\ref{rel}), where now $4/b^2=l\,(l+1)$. In this case, the lowest phonon frequency changes with $b$, that is, $\omega_{ph}=2/b$. By imposing the condition $\omega^2 \ge 0$ for all frequencies, we obtain that the integer number $l^2 \le 4/b^2$. This inequality is satisfied by all integer values of $l$ ($b=2/\sqrt{l\,(l+1)}$). For instance, if $l=1$, then  the value of $b=\sqrt{2}$. If $l=2$, then  the value of $b=\sqrt{2/3}$. As $l$ is increased, the number of discrete modes grows, $b$ decreases, and the stable pulse  becomes broader. 

   \section{Conclusions} \label{sec6}
  
  The stability of  kinks and pulses of the nonlinear Klein-Gordon Eq.\ (\ref{sg}) is investigated by the following procedure: (i)
  It is assumed that its general solution \eqref{sg5} is the superposition of the static solution plus a small perturbation, 
  which depends not only on space, but also on time; (ii) By substituting this ansatz in Eq.\ (\ref{sg}),
  the partial differential equation \eqref{sg6} that governs the perturbation is obtained; (iii) The solution of this 
  equation leads to a Sturm-Liouville problem \eqref{sl-sim}, which is solved in a systematic way 
  for the P\"oschl-Teller potential $-l\,(l+1)\,\sech^{2}(x)$, $l\in\mathbb{N}$.   
   
The detailed resolution of the  Sturm-Liouville problem  \eqref{sl-sim} shows that 
its real eigenvalues are equal to  $\omega^2=\omega_{ph}^2+k^2$ (squared eigenfrequencies), 
where $\omega_{ph}$ is the lowest frequency of the continuous spectrum, and $k^2 \in \mathbb{R}$. 
For $k \ge 0$, we obtain the frequencies of the continuous spectrum $\omega(k)=\sqrt{\omega_{ph}^2+k^2}$. 
Considering $k=i\kappa$, with $\kappa \in \mathbb{R}$, we obtain the frequencies of the discrete spectrum, 
$\omega_{\kappa}=\sqrt{\omega_{ph}^2-\kappa^2}$.  The eigenfunctions of the Sturm-Liouville problem, 
up to a normalizing constant, are $\psi(x)=\exp^{ikx}\,P_l^{(-i k,i k)}[\tanh(x)]$,  
where $P_l^{(-i k,i k)}[\tanh(x)]$ are the  Jacobi polynomials.  Interestingly, 
the degree of the polynomial, $l$,  determines the number of discrete modes, and the parameter $\kappa$ 
takes the values from $1$ to $l$ so that the solution of the Sturm-Liouville problem is bounded. 
  
Furthermore, we establish the orthogonality and completeness relations of this set of eigenfunctions for all values of $l \in \mathbb{N}$.
These results, mentioned in Ref. \cite{rubinstein:1970} for $l=1$ and in Ref. \cite{bishop:1980} for $l=2$, rigorously justify that the solutions of perturbed nonlinear Klein-Gordon equations can be 
written as an expansion in the set of these eigenfunctions.

Starting from the P\"oschl-Teller potential and using the fact that the translational mode is proportional to the spatial derivative of the kink, we obtain a family of nonlinear  Klein-Gordon potentials. Our procedure has two advantages with respect to the previous studies in Refs. \cite{christ:1975,trullinger:1987}. First, our analysis is valid for all values of $l$ and the solution of the second-order differential equation for $U(\varphi)$ is represented in a closed form by Eq.\ \eqref{eqd2} in terms of the hypergeometric function, where $l$ is a parameter. 
	Second, our approach
	shows that the  sine-Gordon and $\varphi^4$ kinks are at the bottom of the hierarchy of stable kinks  
	associated with a certain class of nonlinear Klein-Gordon potentials.

  For the values of $l=1$ and $l=2$, the  spectrum related to  the sine-Gordon and $\varphi^4$ equations, respectively, are recovered \cite{rubinstein:1970,dashen:1974}. 
  Furthermore, we show that, for $l > 2$, there is a family of kinks  corresponding to  Klein-Gordon potentials.  The potential for $l=4$ 
  is obtained explicitly, whereas for $l=3$ and $l \ge 5$, the potentials can be expressed implicitly. 
  Interestingly, we analytically  obtain the kink solutions Eq. \eqref{kinkl} even when the potential can only be numerically found. The  kinks are stable, and are a  
  linear superposition of  two terms: the first is either the sine-Gordon kink  (for $l$ odd numbers) or the $\varphi^4$ kink  (for even $l$), while the second is a localized 
  function. These kinks resemble the $\varphi^4$ wobbling kinks  studied in Ref. \cite{barashenkov:2019}.  The corresponding spectra of the Sturm-Liouville problem associated to the stability of these kinks have several internal modes, some of  which have a localized odd eigenfunction, while others have a localized even eigenfunction.

Finally, we found that if the lowest frequency of the continuous spectrum satisfies $\omega_{ph}<l$ (sufficient condition for instability), then the static solution is unstable.
This is precisely the case of all the studied pulses   $\sech^{n}(x)$ of a family of nonlinear Klein-Gordon equations with a potential given by Eq. \eqref{c1}. 
We explain how certain inhomogeneous terms can be introduced into the nonlinear Klein-Gordon equation in order to obtain stable pulses.  
  
To complete our discussion, the following observations are in order: 
	\begin{enumerate}	
		\item[(1)]  Not all Sturm-Liouville problems associated to the stability problem of the nonlinear Klein-Gordon equation lead to the P\"oschl-Teller 
		potential (see, for instance, Ref. \cite{lohe:1979,barashenkov:1988}). 
		%$\tilde{U}(x)$ for the $\varphi^6$ equ`ation in Table \ref{tabla1}). 
	\item[(2)] Not all the Sturm-Liouville problems associated with the linear stability of static solutions of the nonlinear Klein-Gordon equations  have been analytically solved. 
		For instance, for the double sine-Gordon equation \cite{condat:1983,campbell:1986}, 
		 only   
		its kink solution and the zero mode of its associated Sturm-Liouville problem are known. Indeed, 		
		by taking the spatial derivative of its static kink, it 
 has no zeros. According to our results, all the remaining discrete eigenvalues, if any, are positive, and the double sine-Gordon kink is linearly stable.   However, the computation of the explicit expressions for the remaining eigenfunctions remains an open problem. 
 %are unknown, it is not possible to expand any solution of the perturbed double sine-Gordon by using e the set of eigenfunctions completeness relation  is still being an open problem.
	\end{enumerate}

\appendix

\section{The orthogonality and completeness relations}
 \label{app-close}
In this section,  the orthogonality and completeness relations presented in 
Section \ref{sec-comple} are deduced.
In order to achieve our goal, the theory of the one-dimensional 
Schr\"odinger equation, developed in Chapter 3\S2 of Ref. \cite{takh:2008}, is employed. 
%For doing that we will follow the construction developed in Chapter 3\S2 of Ref. \cite{takh:2008}. 

Instead of dealing with the function $\psi(x,k)$ given by \eqref{sl-cI} and $\psi(x,-k)$, it is convenient 
to use the functions $u_{1}(x,k)$ and $u_{2}(x,k)$, defined below (see Eq. \eqref{continuo-jost-1}). 
First,   two independent solutions of Eq.\ \eqref{sl-sim} are introduced, the so-called 
Jost functions, 
\begin{eqnarray} \nonumber
	f_1(x,k)&=& \frac{\psi(x,k)}{P_l^{(-i k,ik)}(1)}, \\ \label{continuo-jost}
	f_2(x,k)&=&\frac{\psi(x,-k)}{P_l^{(i k,-ik)}(-1)},\quad k>0,
\end{eqnarray} 
with the asymptotics
\begin{eqnarray} \nonumber
	f_1(x,k)&=& e^{ikx}+o(1)\mbox{ as } x\to\infty, \\
	f_2(x,k)&=& e^{-ikx}+o(1)\mbox{ as } x\to-\infty.
\end{eqnarray} 
 Subsequently, using Eq. (2.12) on page 159 of Ref.\cite{takh:2008},  the \textit{transmission} coefficient $a(k)$ is defined,
$$
a(k):=\frac{1}{2ki} W[f_1(x,k),f_2(x,k)]=-\frac{A_{l,k}^2}{P_l^{(-i k,ik)}(1)P_l^{(i k,-ik)}(-1) },
$$
for $k>0$, where $A_{l,k}$ is given by Eq. \eqref{A_k}. 
According to Theorem  2.3 on page 165 in Ref. \cite{takh:2008}, the new functions %have the form 
\begin{eqnarray} \nonumber
	u_\ell(x,k)&=& \frac{f_\ell(x,k)}{a(k)},\quad \ell=1,2,\quad k>0,\\
	u_\ell(x,0)&=& P^{(0,0)}_l(\tanh x),\label{continuo-jost-1}
\end{eqnarray} 
together with the eigenfunctions corresponding to the discrete spectrum \eqref{discrete},  
constitute an orthogonal complete set of functions in $L^2(\mathbb{R})$. 
The orthogonality reads
\begin{equation}
	\label{rel-ort-psi-b}\int_{\mathbb{R}} \psi_\kappa(x) \psi_\nu(x)dx=\delta_{\kappa,\nu},\quad 
	\int_{\mathbb{R}} \psi_\kappa(x)  {u_\ell(x,k)}dx=0,   \quad
\end{equation}
\begin{equation} 
	\label{rel-ort-psi-no-con} \frac{1}{2\pi}
	\int_{\mathbb{R}} \overline{u_\ell(x,k)} u_\iota(x,m)dx=\delta_{\ell,\iota}\delta (k-m),  
\end{equation}
where $\nu,\kappa=1,2\ldots,l$; $\ell,\iota=1,2$; $k,m\geq0$; $\delta_{\kappa,\nu}$ is the Kronecker delta, 
and  $\delta(x)$ denotes the delta Dirac function (which is not actually a function, but a distribution, and hence  Eq. \eqref{rel-ort-psi-no-con} should be understood in the distributional sense). For an 
introduction to the theory of distribution see e.g. Ref. \cite{strichartz:1994}.

On the other hand, for all $\Phi(x)\in L^2(\mathbb{R})$,  one has the expansion 
\cite{takh:2008}. Recall that Eq. (2.25), on page 165 of Ref. \cite{takh:2008}, which is a completeness relation,
is proved for $\Phi\in C_0^2(\mathbb{R})$ (twice continuously differentiable functions on $\mathbb{R}$	with compact 
support). However, since $C_0^\infty(\mathbb{R})$ (infinitely differentiable functions on $\mathbb{R}$ with compact support)  
is a subset of $C_0^2(\mathbb{R})$ and $C_0^\infty(\mathbb{R})$ is dense in $L^2(\mathbb{R})$, then 
$C_0^2(\mathbb{R})$ is dense in $L^2(\mathbb{R})$ and therefore the following expansion is true for all $\Phi\in L^2(\mathbb{R})$
\begin{eqnarray}\label{exp-u-psi}
\Phi(x)= & % \frac{1}{2\pi} 
\displaystyle \frac1{2\pi}\displaystyle \sum_{\ell=1}^2\!\int_{0}^{\infty} \!\!  c_{\ell}(k) u_{\ell}(x,k)dk   
+ \displaystyle \sum_{\kappa=1}^l  c_\kappa 	 \psi_\kappa(x),	
	%	\\ & \nonumber\qquad  + \sum_{\kappa=1}^l   \left(\int_{\mathbb{R}} \psi_\kappa(y)\Phi(y)dy \right)  \psi_\kappa(x),
\end{eqnarray}
where
\begin{eqnarray} \nonumber
	c_\ell(k)&=& \int_{\mathbb{R}} \overline{u_\ell(y,k)}\Phi(y)dy, %\quad \ell=1,2,
	\\ \nonumber
	c_\kappa&=& \int_{\mathbb{R}} \psi_\kappa(y)\Phi(y)dy. %\quad \kappa=1,\ldots,l.
\end{eqnarray}
Formula \eqref{exp-u-psi} is the so-called completeness relation for the set $\{u_1,u_2\}_{k\geq0}
\cup \{\psi_{\kappa}\}_{\kappa=1,\ldots,l}$.
It can also be written in the distributional sense as follows (see Remark on page 168 in Ref. \cite{takh:2008}):
\begin{equation}
	\label{rel-com-psi-dis}% 	\frac{1}{2\pi} 
	\int_{0}^\infty \sum_{\ell=1}^2 \left[  u_{\ell}(x,k)    \overline{u_{\ell}(y,k)} \right] dk + 
	\sum_{\kappa=1}^l  \psi_\kappa(x) \psi_\kappa(y)=\delta (x-y).
\end{equation}
The above orthogonality and completeness relations can be written in a compact form. 
%We will rewrite the above orthogonality and completeness relations in a slightly different way. 
Notice that the integrands of the first terms in Eq. \eqref{exp-u-psi}  are 
$$
c_\ell(k) u_\ell(x,k) = \left(\int_{0}^\infty  \overline{u_\ell(y,k)}\Phi(y)dy \right) u_\ell(x,k),\qquad \ell=1,2.
$$
Using $|P_l^{(i k,-ik)}(-1)|^2=|P_l^{(-i k,ik)}(1)|^2=A_{l,k}^2$, it is straightforward to deduce  that  
\begin{eqnarray*}
	\overline{u_1(y,k)} u_1(x,k)&=& \frac{\overline{\psi(y,k)}\psi(x,k)}{A_{l,k}^2}, \\ 
	\overline{u_2(y,k)} u_2(x,k)&=& \frac{\overline{\psi(y,-k)}\psi(x,-k)}{A_{l,k}^2}.
\end{eqnarray*}
Using the above identities and changing $k\to -k$ in the second integral of Eq. \eqref{exp-u-psi}, 
 this expression becomes 
\begin{eqnarray*}%\label{exp-psi}
	\Phi(x)= & \displaystyle \frac{1}{2\pi} 
	\int_{\mathbb{R}} \left[ A_{l,k}^{-2} \left(\int_{\mathbb{R}} \overline{\psi(y,k)}\Phi(y)dy \right) \psi(x,k) \right]dk 
	\\ & \nonumber\qquad \displaystyle + \sum_{\kappa=1}^l   \left(\int_{\mathbb{R}} \psi_\kappa(y)\Phi(y)dy \right)  \psi_\kappa(x).
\end{eqnarray*}
From the above equation it follows that the set of functions defined by Eqs. \eqref{discrete}-\eqref{continuo-equiv}
 satisfies the completeness relation  \eqref{exp-psi} and its equivalent expression
\eqref{rel-com-psi}. In a similar way the orthogonality relations 
\eqref{rel-ort-psi-b}-\eqref{rel-ort-psi-no-con} become 
the relations \eqref{rel-ort-psi-b-equ}-\eqref{rel-ort-psi}, respectively. 

The set of functions defined by Eqs. \eqref{discrete} and \eqref{continuo-equiv} that satisfy the relations 
\eqref{rel-com-psi}, \eqref{rel-ort-psi-b-equ}, and \eqref{rel-ort-psi} are those  
 used in the theory of the nonlinear Klein-Gordon Eq. \eqref{sg}.

\section*{Acknowledgments}
R.A.N. was partially supported by PGC2018-096504-B-C31 (FEDER(EU)/ Ministerio de Ciencia e Innovaci\'on-Agencia Estatal de Investigaci\'on), 
FQM-262 and Feder-US-1254600 (FEDET(EU)-Jun\-ta de Anda\-lu\-c\'ia).
N.R.Q. was partially supported by the Spanish projects PID2020-113390GB-I00 (MICIN), PY20$\_$00082 
(Junta de Andalucia),  A-FQM-52-UGR20 (ERDF-University of Granada), and the Andalusian research group FQM-207.
 
%%%%%%%%%%%%%%%%%%%%%%%%%%%%%%%%%%%%%%%%%%%%%%%%%%%%%%%%%%%%%%%%%%%%%%%%%%

\end{document}